\documentclass[twocolumn,showpacs,preprintnumbers,amsmath,amssymb,aps]{revtex4-1}
\usepackage{graphicx}
\usepackage{dcolumn}
\usepackage{bm}

\begin{document}

\title{Spin dynamics and magnetic-field-induced polarization of excitons in
ultrathin GaAs/AlAs quantum wells with indirect band gap and type-II band alignment}

\author{T.~S.~Shamirzaev$^{1,4}$,  J. Rautert$^2$, D.~R.~Yakovlev$^{2,5}$, J.~Debus$^2$, A.~Yu.~Gornov$^3$, M.~M.~Glazov$^{5}$, E.~L.~Ivchenko$^{5}$, and M.~Bayer$^{2,5}$}

\affiliation{
$^1$Rzhanov Institute of Semiconductor Physics, Siberian Branch of the Russian Academy of Sciences, 630090 Novosibirsk, Russia \\
$^2$Experimentelle Physik 2, Technische Universit\"at Dortmund, 44227 Dortmund, Germany \\
$^3$Institute for System Dynamics and Control Theory, Siberian Branch of the Russian Academy of Sciences, 664033 Irkutsk, Russia \\
$^4$Ural Federal University, 620002 Yekaterinburg, Russia \\
$^5$Ioffe Institute, Russian Academy of Sciences, 194021 St. Petersburg, Russia }

\begin{abstract}
The exciton spin dynamics are investigated both experimentally and theoretically in two-monolayer-thick GaAs/AlAs quantum wells with an indirect band gap  and a type-II band alignment. The magnetic-field-induced circular polarization of photoluminescence, $P_c$, is studied as function of the magnetic field strength and  direction as well as sample temperature. The observed nonmonotonic behaviour of these functions is provided by the interplay of bright and dark exciton states contributing to the emission. To interpret the experiment, we have developed a kinetic master equation model which accounts for the dynamics of the spin states in this exciton quartet, radiative and nonradiative recombination processes, and redistribution of excitons between these states as result of spin relaxation. The model offers quantitative agreement with experiment and allows us to evaluate, for the studied structure, the heavy-hole $g$ factor, $g_{hh}=+3.5$, and the spin relaxation times of electron, $\tau_{se} = 33~\mu$s, and hole, $\tau_{sh} = 3~\mu$s, bound in the exciton.
\end{abstract}


\maketitle
\section{Introduction}
\label{sec:intro}

Heterostructures with semiconductor quantum wells (QWs) have been interesting
from several points of view including basic physics and optoelectronic
devices~\cite{excitonbook, Klingshirn}. Among the different types of QWs, the
longest exciton lifetimes up to milliseconds are obtained in indirect band-gap heterostructures with type-II band alignment~\cite{Fu}.
Here the long lifetime is due to the separation of the oppositely charged carriers forming the excitons in real and momentum space. Since
the electron spin relaxation times
may reach milliseconds according to theoretical estimations~\cite{Khaetskii0, Khaetskii}, type-II
heterostructures, such as GaAs/AlAs QWs and superlattices, are highly
interesting for studying long-lived exciton spin dynamics, which are not limited by
exciton recombination. Recently, we demonstrated that an external magnetic field can  control the
intensity of the long-lived exciton recombination in type-II GaAs/AlAs
QWs via redistribution of exciton population between bright and
dark states~\cite{Shamirzaev3}. However, the magnetic-field-induced polarization
and the related spin dynamics of excitons in such QWs have been scarcely
studied so far.

In this paper, we investigate the effects of a magnetic field on the
exciton spin dynamics in ultrathin GaAs/AlAs QWs with indirect band
gap and type-II band alignment. The circular polarization degree ($P_c$)
of the QW exciton photoluminescence (PL) induced by a magnetic field
shows an unusual behavior: (i) $P_c$ has a strong spectral dependence
across the emission band of the QW; it is small for the no-phonon (NP) line but its absolute value increases strongly for lines of phonon-assisted recombination.
(ii) In tilted magnetic field $|P_c|$ demonstrates a monotonic
increase with saturation in high fields, whereas in the Faraday or close-to-Faraday geometry
$P_c$ shows a non-monotonic behavior as function of the magnetic field $B$. Namely, the modulus of the polarization degree increases in low magnetic fields, reaches a maximum, and then
decreases in strong fields. These experimental appearances are
explained in the framework of a theoretical model based on the approach developed in our previous work~\cite{Shamirzaev3}.

The paper is organized as follows. In Sec.~\ref{sec:1} the studied sample
and used experimental techniques are described. In Sec.~\ref{sec:2}
we present the experimental data on the $P_c$ of time-integrated and
time-resolved PL recorded in external magnetic fields. The kinetic
equation model, which accounts for the dynamics of the quartet of bright
and dark exciton states in ultrathin GaAs/AlAs QWs, is presented
in Sec.~\ref{sec:theory}. The experimental data are analyzed in the
frame of the model in Sec.~\ref{sec:discussion}, which allows us to
evaluate the carrier spin relaxation times and $g$ factors. We show that
the experimental data, being controlled by
these parameters, can vary strongly, and we perform model calculations for several cases of interest for
future experimental studies.

\section{Experimental details}
\label{sec:1}

The ultrathin GaAs/AlAs QW structure studied here was grown by molecular-beam
epitaxy on a semi-insulating (001)-oriented GaAs substrate in a
Riber Compact system. The sample consists of the GaAs QW layer
embedded between 50-nm-thick layers of AlAs grown on top of a
200-nm-thick GaAs buffer layer~\cite{choiceofGaAs}. The substrate
temperature during the growth was 600$^\circ$C. The GaAs QW layer
was deposited with a nominal thickness of two monolayers. A
20-nm-thick GaAs cap layer protects the top AlAs layer against
oxidation. Further growth details are given in
Ref.~\cite{Shamirzaev1}. The GaAs/AlAs QW has a type-II band
alignment with the lowest conduction-band states at the
X$_{x}$ and X$_y$ minima of the AlAs conduction
band~\cite{Shamirzaev1,Fu,Poel}. A schematic band diagram of the
structure and the optical transition of the indirect exciton to the
system ground state are presented in the inset of Fig.~\ref{fig1}(b).

The sample was placed in a split-coil magnet cryostat and exposed to
magnetic fields up to $B=10$~T. The angle $\theta$ between the
magnetic field direction and the QW growth axis ($z$ axis) was
varied between 0$^\circ$ (Faraday geometry) and 90$^\circ$ (Voigt
geometry). For measurement of angular dependencies we
fixed the magnetic field direction and rotated the sample. The emission
was collected either in the direction along the field direction in Faraday geometry for 0$^\circ~$
$\le~\theta~\le$~45$^\circ$ or perpendicular to the field direction in Voigt geometry for 45$^\circ~< \theta~\le$~90$^\circ$.
The temperature was varied from $T=1.8$~K up to 19~K.
The photoluminescence  was excited by the third harmonic of a
Q-switched Nd:YVO$_{4}$ laser (3.49~eV) with a pulse duration of
5~ns. The pulse energy density was kept below $100$~nJ/cm$^2$ and
the pulse-repetition frequency was varied from 20~Hz up to
1~kHz~\cite{Shamirzaev3}.

The emitted light was dispersed by a 0.5-m
monochromator. For time-integrated measurements the
photoluminescence was detected by a liquid-nitrogen-cooled
charge-coupled-device (CCD) camera.
For the time-resolved measurements a GaAs photomultiplier combined
with a time-correlated photon-counting module was used. In order to
monitor the PL decay in a wide temporal range of up to 30~ms, the
time resolution of the detection system (i.e. the binning range of
the photon counting events) was varied between 3.2~ns
and 6.4~$\mu$s.

The exciton spin dynamics were analyzed from the PL by measuring the
circular polarization degree $P_c$ induced by the external magnetic
field. $P_c$ was evaluated from the data by $P_c =
[I_{\sigma^+}-I_{\sigma^-}]/[I_{\sigma^+}+ I_{\sigma^-}]$, where
$I_{\sigma^+}$ and $I_{\sigma^-}$ are the intensities of the
$\sigma^{+}$ and $\sigma^{-}$ polarized PL components, respectively.
To determine the sign of $P_c$, we performed a control measurement
on a diluted magnetic semiconductor structure with
(Zn,Mn)Se/(Zn,Be)Se quantum wells for which $P_c>0$ in Faraday
geometry~\cite{Keller}.

\section{Experimental results}
\label{sec:2}

\begin{figure}[hbt]
\includegraphics* [width=7.5cm]{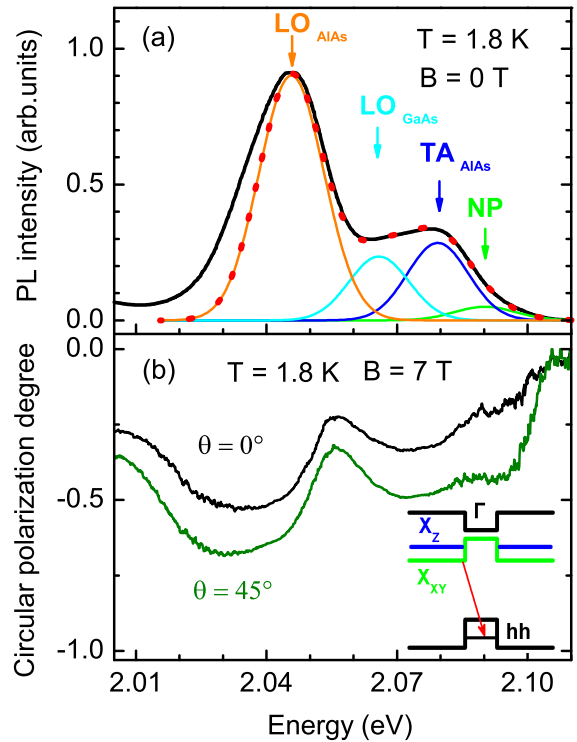}
\caption{\label{fig1} (Color online)  (a) Time-integrated PL
spectrum of the ultrathin GaAs/AlAs QW (black line) fitted with four
Gaussian lines corresponding to the exciton recombination with and
without involvement of phonons. Green, blue, cyan, and orange solid
lines are the no-phonon, TA$_{\mathrm {AlAs}}$, LO$_{\mathrm
{GaAs}}$ and LO$_{\mathrm {AlAs}}$ phonon lines, respectively. The red
dotted line is the fitted spectrum composed of the four lines. (b)
Spectral dependence of the PL circular polarization degree induced by
a longitudinal ($\theta$ = 0$^\circ$) and a tilted ($\theta$ =
45$^\circ$) magnetic field $B$ = 7~T. The inset shows the schematic band
alignment of the structure with the QW in the middle. The red arrow marks the optical
transition of the indirect exciton to the system ground state.}
\end{figure}

A time-integrated photoluminescence spectrum of the ultrathin
GaAs/AlAs QW is shown in Fig.~\ref{fig1}(a) by the black line. The
spectrum comprises the contributions from several
emission processes, as it contains the no-phonon (NP) line and several lines
of phonon-assisted recombination involving optical and acoustic
phonons from  GaAs and AlAs. The replicas associated with the
transverse acoustic (TA) phonons of AlAs (phonon energy of 12~meV) and the
longitudinal optical (LO) phonons of GaAs (30~meV) and AlAs
(48~meV), all at the $X$ point  of the Brillouin
zone~\cite{NumeralData}, can be distinguished. The lines are broadened
due to the roughness of the QW interfaces~\cite{Shamirzaev1}. An
example of fitting of the PL spectrum with four
contributing Gaussian curves, each with the same width of 19~meV, is
shown in Fig.~\ref{fig1}(a) by the red dotted line~\cite{Shamirzaev3}.

Application of a magnetic field results in polarization of the
emission, as shown in Fig.~\ref{fig1}(b) for $B$ = 7~T in the
Faraday geometry. One can see that $P_c$ is negative
(i.e., it isdominated by the $\sigma^{-}$ polarized PL component) and has
a strong spectral dependence. The absolute value of $P_c$ equals to 0.17 for the NP line and increases up
to 0.33 and 0.53 for the TA$_{\mathrm {AlAs}}$ and LO$_{\mathrm {AlAs}}$
phonon-assisted lines, respectively.

\begin{figure}[hbt]
\includegraphics* [width=7.0cm]{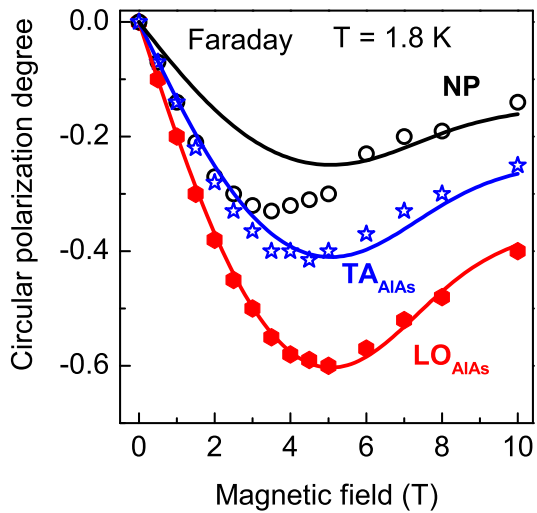}
\caption{\label{fig2} (Color online) Circular polarization degree
induced by a longitudinal magnetic field as function of field
strength for the NP (open circles), TA$_{\mathrm {AlAs}}$ (open stars)
and LO$_{\mathrm {AlAs}}$ (full hexagons) lines, respectively. Lines
show results of modeling with parameters given in Sec.~\ref{subsec:modelling}.}
\end{figure}

Next we take a closer look at the longitudinal magnetic field effect
on the no-phonon and phonon-assisted lines. Since the intensities of
the $\sigma^+$ and $\sigma^-$ polarized PL components are
proportional to the populations of the exciton Zeeman sublevels,
$P_c$ reflects these populations. The $P_c(B)$ dependencies of these
lines measured in longitudinal magnetic field are shown in
Fig.~\ref{fig2}. For all lines the polarization degree increases in
low magnetic fields, reaches a maximum, and then decreases in strong
fields. The maximal value of the polarization degree, $|P_{c,max}|$,
is achieved at magnetic fields of $B_{max}= 3.7$~T, 4.6~T, and 5~T
for the NP, TA$_{\mathrm {AlAs}}$ and LO$_{\mathrm {AlAs}}$ lines,
respectively. While the difference of $B_{max}$ for the TA and LO
phonon replicas is negligible, the NP line strongly deviates from
the phonon replicas. The origin of this line is not well
established~\cite{Shamirzaev3}: In particular, it may be related
with trions, i.e., negatively charged excitons. In that case, the
selection rules are strongly different from those for neutral
excitons. Therefore, we exclude this line from the analysis. Note
that all replicas show similar changes in intensity,
dynamics~\cite{Shamirzaev3} and polarization with varying
temperature, magnetic field strength and sample orientation.
Therefore, we will focus on the properties of the phonon-assisted
LO$_{\mathrm {AlAs}}$ line.

\begin{figure}[hbt]
\includegraphics* [width=7.0 cm]{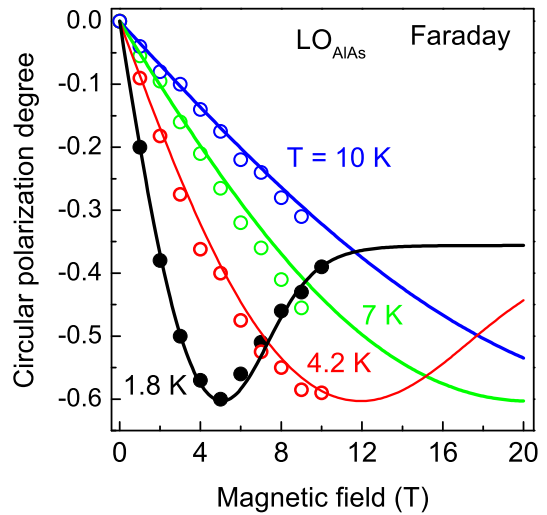}
\caption{\label{fig3} (Color online) Magnetic-field-induced circular polarization degree of
LO$_{\mathrm {AlAs}}$ phonon-assisted transition
measured at temperatures of 1.8, 4.2, 7 and 10~K (symbols). Lines
show results of modeling with parameters given in Sec.~\ref{subsec:modelling}.}
\end{figure}

\begin{figure}[hbt]
\includegraphics* [width=7.0 cm]{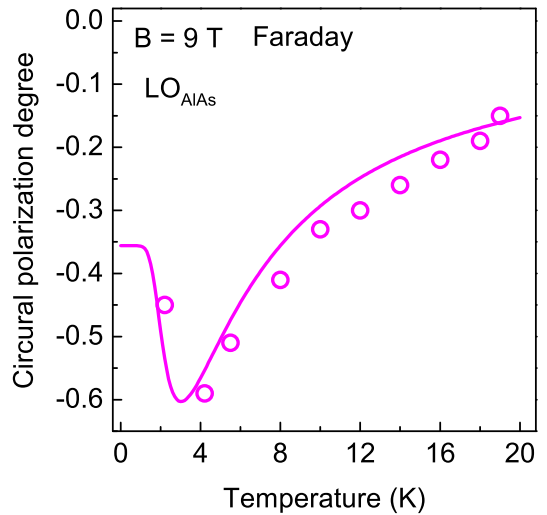}
\caption{\label{fig4} (Color online) Magnetic-field-induced circular polarization degree of the
LO$_{\mathrm {AlAs}}$ phonon-assisted transition at $B=9$~T as function of temperature. Circles give
experimental data and line is the modeling result with the parameters given in Sec.~\ref{subsec:modelling}.}
\end{figure}

The $P_c(B)$ dependencies for the LO$_{\mathrm {AlAs}}$ line at
various temperatures are shown in Fig.~\ref{fig3}. With increasing
temperature the slope of the polarization rise monotonically
decreases and the $P_{c,max}$ value is shifted towards stronger
magnetic fields. For a fixed longitudinal magnetic field of 9~T,
$P_c(T)$ demonstrates an unexpected nonmonotonic dependence as shown
in Fig.~\ref{fig4}. The absolute value of the polarization degree
increases from 0.45 up to 0.59 with increasing temperature from 2 up
to 4.2~K. Then, it steadily decreases down to 0.15 with further
temperature increase to 19~K.

\begin{figure}
\includegraphics* [width=7cm]{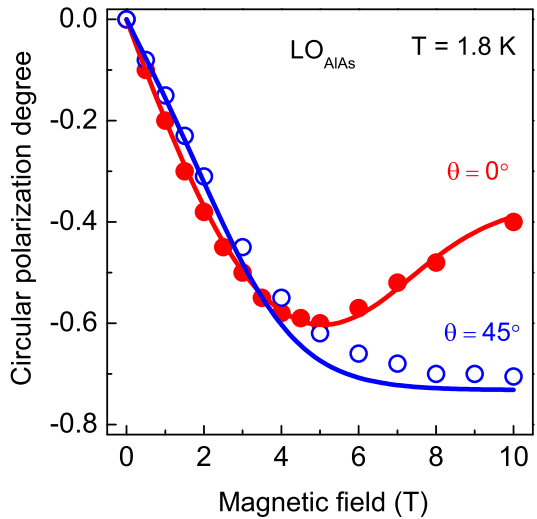}
\caption{\label{fig5} (Color online)  Magnetic-field-induced circular polarization degree
of the LO$_{\mathrm {AlAs}}$ phonon-assisted transition, in longitudinal (closed red circles) and tilted by
45$^\circ$ (open blue circles) field configuration as function of the field strength.
Lines are the modeling results with the parameters given in Sec.~\ref{subsec:modelling}.}
\end{figure}

The $P_c(B)$ of the phonon-assisted lines depends strongly on the
experimental geometry as shown in Fig.~\ref{fig5}. It behaves
nonmonotonically in the Faraday geometry, while a finite angle
$\theta$ between the magnetic field direction and the QW growth axis
unexpectedly suppresses the decrease of $|P_c|$ in high magnetic
fields. $|P_c(B)|$  for a field tilt by 45$^\circ$ from the Faraday
geometry increases monotonically and saturates at 0.72 for 10~T.
However, as one can see in Fig.~\ref{fig1}(b), the spectral
dependence of the polarization degree in the tilted geometry is
pronounced and similar to that in the Faraday geometry.

\begin{figure}[hbt]
\includegraphics* [width=\linewidth]{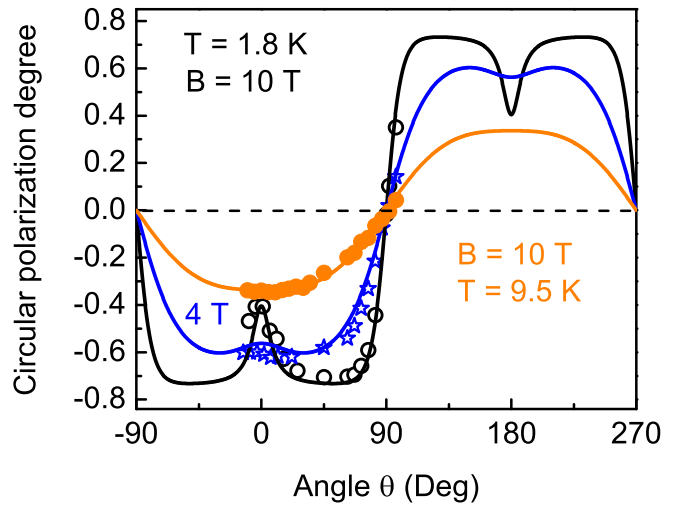}
\caption{\label{fig6}(Color online) Angle dependencies of $P_c(\theta)$
measured at temperature $T=1.8$~K for $B=4$~T (open blue stars) and 10~T (open
black circles), and at $T=9.5$~K for $B=10$~T (full
orange circles). Lines show results of the modeling with the parameters
given in text.}
\end{figure}

In order to analyze in more detail the angular dependence of the
polarization degree, we measured it at $T=1.8$~K for $B=4$ and 10~T,
and at $T=9.5$~K for $B=10$~T. These results are shown in
Fig.~\ref{fig6}.  For low magnetic fields, where $|P_c|$ is smaller
than its maximal value $|P_{c,max}|$ in longitudinal magnetic field,
the polarization degree changes weakly up to about $\theta =
45^\circ$, and goes to zero for $\theta$ increased further to
90$^\circ$ (Voigt geometry). In higher magnetic fields $P_c$
increases for 0$^\circ < \theta < $ 30$^\circ$, changes
insignificantly in the range of 30$^\circ <  \theta <$ 60$^\circ$,
and goes to 0 for $\theta$ enlarged towards 90$^\circ$. For any
magnetic field an increase of $\theta$ beyond 90$^\circ$ results in
a change of the polarization sign; the PL is dominated then by the
$\sigma^{+}$ polarized component. With increasing temperature up to
9.5~K at $B=10$~T, the polarization ${|P_c(\theta)|}$ monotonically
decreases when $\theta$ is varied from 0 to 90$^\circ$.

\begin{figure}[hbt]
\includegraphics* [width=7 cm]{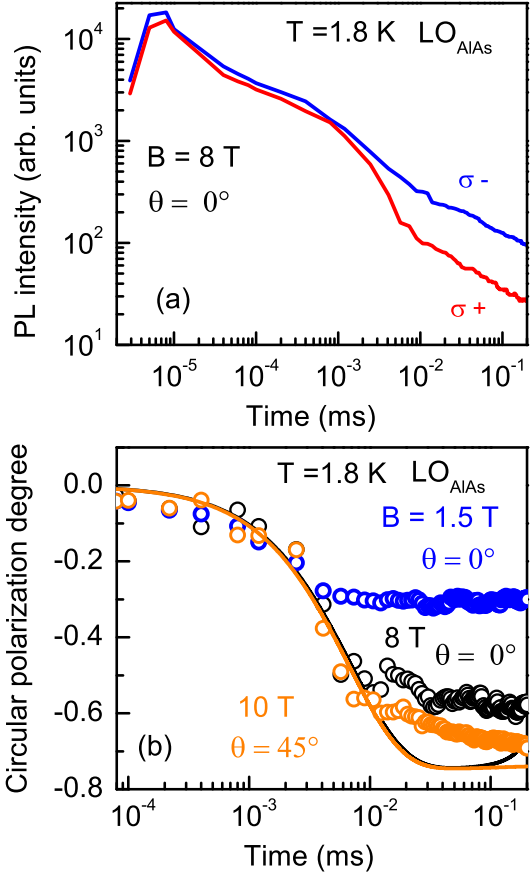}
\caption{\label{fig7} (Color online) (a) Recombination dynamics of
the ${\sigma^{+}}$ and ${\sigma^{-}}$ polarized PL components of the
LO$_{\mathrm {AlAs}}$ emission line, measured at $T = 1.8$~K in the
Faraday geometry for $B= 8$~T. (b) Dynamics of the circular
polarization degree of the LO$_{\mathrm {AlAs}}$ emission line
measured at $T = 1.8$~K for different magnetic field strengths and
geometries: $B = 1.5$~T (blue symbols) and 8~T (black symbols) in
the Faraday geometry, and $B = 10$~T (orange symbols) tilted by
45$^\circ$. Solid lines show the model results for $B = 8$~T (black
line) in the Faraday geometry and $B = 10$~T (orange line) for
$\theta = 45^\circ$ with the parameters given in text. The
calculations of $P_c(t)$ dynamics are made via Eq.~\eqref{Pc0} using
the solution of Eqs.~\eqref{anyB} for pulsed excitation. The
parameters are given in Sec. V A.}
\end{figure}

The PL dynamics for different magnetic field strengths and
geometries for the two circular polarized components, i.e.
$I_{\sigma^+}(t)$ and $I_{\sigma^-}(t)$, and the dynamics of
the circular polarization degree
$P_c(t)=[I_{\sigma^+}(t)-I_{\sigma^-}(t)]/[I_{\sigma^+}(t)+I_{\sigma^-}(t)]$
are shown in Fig.~\ref{fig7}. The circular polarization is very small
for short delay times, then it increases within a characteristic time of
${\tau_s}=3~\mu$s, which is much shorter than the exciton
PL decay time~\cite{Shamirzaev3}. It is interesting that
the change in magnetic field strength and geometry results
in a modification of the maximal value of the polarization degree only, while
${\tau_s}$ remains basiclly constant in the whole scanned $B$ and $\theta$ range.

In conclusion of this section, we summarize the most important
experimental findings:
\begin{itemize}
\item[(i)] The magnetic-field-induced circular
polarization degree $P_c$ has a complicated spectral
dependence.
\item[(ii)] In the Faraday (and close-to-Faraday) geometry the absolute value of $|P_c|$
increases in low magnetic fields and decreases in strong
fields. However, in tilted magnetic field, $|P_c(B)|$
demonstrates a monotonic increase with saturation in strong fields.
\item[(iii)] In the Voigt geometry the circular polarization degree
$P_c=0$. Varying $\theta$ from 90${^\circ}$ by a few degrees leads to a rapid increase in $P_c$.
\item[(iv)] The temperature dependence of $P_c$ is also non-monotonic. In
high magnetic fields the absolute value of the polarization degree
increases with increasing temperature from 2 up to 4.2~K, and then it
monotonically decreases with further temperature increase.
\end{itemize}

\section{Theory}
\label{sec:theory}

In this section we present a kinetic theory of the
photoluminescence polarization in monolayer-thin GaAs/AlAs quantum
wells. In Sec.~\ref{subsec:kin} we introduce the kinetic equations
for the occupancies of the quadruplet of exciton spin sublevels. In Sec.~\ref{subsec:an} we present the analytical solution of these kinetic equations for arbitrary strength and direction of the
magnetic field. We analyze various limiting cases of the Faraday and
Voigt geometries in Secs.~\ref{subsec:F} and \ref{subsec:V},
respectively, and address the selection rules in
Sec.~\ref{subsec:selection}.

\subsection{Kinetic equations for the occupation probabilities of exciton spin sublevels}
\label{subsec:kin}

We consider the quadruplet of exciton states formed by a
conduction-band electron and a heavy hole in the valence band.
Following the model of Ref.~\cite{Shamirzaev3} we neglect the
exchange interaction between the electron and the hole and assume
that the electron Zeeman effect in $X_{x,y}$-valleys is isotropic,
i.e., the in- and out-of-plane components of the $g$ factor are
equal. By contrast, we completely disregard the in-plane magnetic
field effect on the heavy-hole spin. As a result, the eigenstates of
the hole are still characterized by the $z$-component of the angular
momentum, $j_z=\pm 3/2$, and denoted as $|\pm 3/2\rangle_z$. The
electron eigenstates $|s\rangle_{\bm B}$ are characterized by the
spin component $s=\pm 1/2$ onto the magnetic field direction and
form the superpositions of the basic functions $|\pm 1/2\rangle_z$:
\begin{align} \label{Bz}
&| +1/2 \rangle_{\bm B} = ~~C \vert 1/2  \rangle_z + D \vert -1/2  \rangle_z \:,\\
&|  -1/2 \rangle_{\bm B} = - D \vert 1/2 \rangle_z  + C \vert - 1/2 \rangle_z  \:. \nonumber
\end{align}
Here, the subscripts $z$ and ${\bm B}$ are introduced in order to
highlight the difference between the basic spinors and the
eigenfunctions of the Zeeman Hamiltonian, the coefficients $C$ and
$D$ (normalized to unity) depend on the magnetic field orientation.
If the magnetic field is tilted by the angle $\theta$ with respect
to the sample normal in the $(xz)$ plane the coefficients in
Eq.~\eqref{Bz} take the simple form:
\begin{equation}
\label{CD:coeffs}
C = \cos{(\theta/2)}, \quad D = \sin{(\theta/2)}.
\end{equation}
Hence, the exciton spin state is represented as a product of the electron and the heavy-hole eigenstates
\[
|s j_z\rangle = |s\rangle_{\bm B} |j_z\rangle_z,
\]
and labeled by the pair of electron and hole spin components, $s$
and $j_z$. Within the model presented in
Ref.~\cite{Shamirzaev3}, which accounts for spin flips of
the electron and hole as well as the radiative and nonradiative
recombination of the exciton, the occupancies, $f_{sj_z}$, of the
exciton states  obey the following set of kinetic equations:
\begin{eqnarray}
\label{kinetic}
\frac{d f_{sj_z}}{dt} &+& \left( W_{\bar s, s} + W_{\bar j_z, j_z} \right) f_{sj_z} - W_{s,\bar s} f_{\bar sj_z}  - W_{j_z,\bar j_z} f_{s\bar j_z}\nonumber \\
 &+& \mathcal R f_{sj_z} = G_{{sj_z}}\:.
\end{eqnarray}
Here $\bar s = -s$, $\bar j_z = -j_z$, $W_{s,s'}$ ($W_{j_z,j_z'}$)
are the electron (heavy-hole) spin-flip rates for the transitions
$s' \to s$ ($j_z' \to j_z$), the operator $\mathcal R$ describes the
radiative and nonradiative recombination of excitons, and
$G_{{sj_z}}$ is the exciton generation rate in the state
$|sj_z\rangle$. In accordance with Ref.~\cite{Shamirzaev3} we
present the energies of the exciton sublevels $E_{sj_z}$ in the form
\begin{equation}
\label{energies}
E_{sj_z} = g_e s \mu_B B +\frac{g_{hh}}{3}j_z\mu_B B_z,
\end{equation}
where $\mu_B$ is the Bohr magneton, $g_e$ and $g_{hh}$ are the
electron and heavy-hole Land\'{e} factors, $B$ is the total magnetic
field, $B_z$ is its $z$ component. In the experiment the Zeeman
splittings can be comparable with the thermal energy, $k_{\rm B} T$,
where $T$ is the sample temperature and $k_{\rm B}$ is the Boltzmann
constant. Hence, the rates of the transitions from the lower to
higher and from the higher to lower Zeeman sublevels are different
and can be interrelated as
\begin{eqnarray}
&&W_{1/2,-1/2} = W_{-1/2,1/2} \exp{\left(-\frac{g_e \mu_B B}{k_{B} T}\right)},\label{rates}\\
&&W_{3/2,-3/2}= W_{-3/2,3/2} \exp{\left(-\frac{g_{hh} \mu_B
B_z}{k_{B} T}\right)}\:.\nonumber
\end{eqnarray}
Both electron and hole Land\'{e} factors are positive in the studied
sample, moreover, $g_{hh} > g_e$~\cite{Shamirzaev3}. Then, without
loss of generality, we choose the positive direction of the $z$ axis
to have $B_z > 0$ and represent the transition rates in the form
\begin{eqnarray} \label{wewh}
&&W_{-1/2,1/2} \equiv w_e, \quad W_{-3/2,3/2} \equiv w_h, \\
&&W_{1/2,-1/2} = \alpha w_e, \quad W_{3/2,-3/2} = \beta w_h, \nonumber
\end{eqnarray}
where the Boltzmann factors in Eqs.~\eqref{rates} are denoted as
\begin{equation}
\label{Boltz} \alpha = \exp{\left(-\frac{g_e \mu_B B}{k_{B}
T}\right)}, \quad \beta = \exp{\left(-\frac{g_{hh} \mu_B B_z}{k_{B}
T}\right)},
\end{equation}
and $\alpha,\beta\leqslant 1$. The rates $w_e$, $w_h$ represent the spin-flip transitions downwards in energy, their dependence on magnetic field is relatively weak and described by power laws~\cite{Khaetskii,Woods,Xiayu} and is disregarded hereafter. In the following we simplify the
notation for the occupation probability and replace $f_{sj_z}$
$(s=\pm 1/2, j_z = \pm 3/2)$ just by $f_{ij}$ with $i=\pm$ and
$j=\pm$ indicating the signs of $s$ and $j_z$, respectively.

Let us now establish the selection rules governing the
phonon-assisted exciton recombination in ultrathin
GaAs/AlAs quantum wells, characterized by an indirect band gap and a
type-II band alignment. It is commonly accepted that the recombination of the
exciton in such system is related with (i) the spin-conserving virtual transition of the electron from the $X_x$
or $X_y$ valley to the $\Gamma$-point accompanied by phonon
emission and (ii) the recombination of the exciton at the Brillouin
zone center~\cite{Xxy2,Dawson,Shamirzaev3}. At the $\Gamma$-point
the strict selection rules involving the periodic Bloch amplitudes
of the $|\Gamma_6, s_z\rangle$ conduction band electron state and
the $|\Gamma_8,j_z\rangle$ valence band hole state characterized by
their spin $z$-components ${s_z}, j_z$ read

\begin{eqnarray} \label{1}
&&|\Gamma_6,1/2; \Gamma_8,-3/2 \rangle \to \sigma_-\:,\: |\Gamma_6,-1/2;\Gamma_8, 3/2 \rangle \to \sigma_+,\\
&&|\Gamma_6,1/2;\Gamma_8, 1/2 \rangle \to \sigma_+\:,\: |\Gamma_6,-1/2;\Gamma_8, -1/2 \rangle \to \sigma_-\:, \nonumber\\
&&  |\Gamma_6,1/2;\Gamma_8,3/2 \rangle, \: |\Gamma_6,-1/2; \Gamma_8,-3/2 \rangle \to {\rm forbidden}\:,\nonumber \\
&& |\Gamma_6,1/2;\Gamma_8, -1/2 \rangle, |\Gamma_6,-1/2;\Gamma_8, 1/2 \rangle \to z\mbox{-polarized}\:. \nonumber
\end{eqnarray}

Neglecting the mixing of the hole states we obtain that only the exciton states with $s_z= - 1/2, j_z = 3/2$ and $s_z= 1/2, j_z = -3/2$ are active (bright) in $\sigma^+$ and $\sigma^-$ polarization~\cite{ivchenko05a,Shamirzaev3}, respectively. These selection rules result in the following form of the recombination operator $\mathcal R$:
\begin{align}\label{recomb}
\mathcal R f_{-+/+- }  = \left( \frac{1}{\tau_{nr}} + \frac{C^2}{\tau_{r}} \right) f_{-+/+- },\\
\mathcal R f_{++/--}  = \left(  \frac{1}{\tau_{nr}} + \frac{D^2}{\tau_{r}} \right) f_{++/--}, \nonumber
\end{align}
where the subscript $-+$/$+-$ means $-+$ {\sl or} $+-$, $\tau_{nr}$ and
$\tau_r$ are the nonradiative and radiative recombination times of the
excitons, and the coefficients $C,D$ are defined by
Eq.~(\ref{CD:coeffs}). The heavy-hole exciton emits light which
propagates along the quantum well normal. Correspondingly, the
intensities of the exciton emission in $\sigma^+$ and $\sigma^-$
polarization are given by
\begin{equation}
\label{intens}
I_{+/-} \propto \frac{C^2}{\tau_{r}} f_{-+/+-} + \frac{D^2}{\tau_{r}} f_{++/--}\:.
\end{equation}
Therefore, one obtains for the degree of circular polarization of the emission
\begin{equation}
\label{Pc0}
P_c = \frac{C^2 (f_{-+}-f_{+-}) +D^2 (f_{++}-f_{--})}{C^2 (f_{-+}+f_{+-}) + D^2 (f_{++}+f_{--})}.
\end{equation}
Deviations from these strict selection rules are addressed in Sec.~\ref{subsec:selection}.

For arbitrary orientation of the external  magnetic field
in the interval $0 \leq \theta \leq \pi/2$, we get from
Eqs.~\eqref{kinetic}, \eqref{wewh} and \eqref{recomb}:
\begin{widetext}
\begin{eqnarray} \label{anyB}
&& \frac{d f_{--}}{dt} + \left(D^2 w + w' \right) f_{--} + w_e (\alpha f_{--} - f_{+-})  + w_h (\beta f_{--} - f_{-+}) = G_{--}\:,\\
&&  \frac{d f_{++}}{dt} + \left(C^2 w + w' \right) f_{++} + w_e (f_{++} - \alpha f_{-+})  + w_h (f_{++} - \beta  f_{+-}) = G_{++}\:, \nonumber\\
&& \frac{d f_{+-}}{dt} + \left(w' + C^2w\right) f_{+-} + w_e (f_{+-} - \alpha  f_{--})  + w_h (\beta f_{+-} - f_{++}) = G_{+-}\:, \nonumber\\
&& \frac{d f_{-+}}{dt} + \left(w' + D^2 w \right) f_{-+} + w_e (\alpha f_{-+} - f_{++})  + w_h (f_{-+} - \beta  f_{--}) = G_{-+} \:,\nonumber
\end{eqnarray}
\end{widetext}
where $w=1/\tau_{r}$ and $w'=1/\tau_{nr}$. Note that for $\theta=0$,
i.e., for the Faraday geometry, a similar set of kinetic equations
describing bright and dark exciton spin dynamics was presented in
Ref.~\cite{rodina2014}. This model is sufficient to describe the
recombination dynamics of excitons observed in
Ref.~\cite{Shamirzaev3} and, as shown below, readily accounts for
most of the present experimental observations.

\subsection{Analytical solution of the set (\ref{anyB}) in the steady-state regime}
\label{subsec:an}

Under steady-state photoexcitation the time derivatives in Eqs.
(\ref{anyB}) vanish, $df_{sj_z}/dt=0$. Here and in what follows we
focus on the important limit of equal generation rates in all exciton
states, $G_{ij} \equiv G$, which is relevant for the experimental
situation of interest. Making use of the firs t two equations of the
set~\eqref{anyB} one can express the occupancies of the (in the Faraday geometry) dark
states, $f_{++}$ and $f_{--}$
through the occupancies of the bright states $f_{-+}, f_{+-}$ as follows
\begin{eqnarray}
\label{f2}
f_{--}&=& \frac{G + w_e f_{+-} + w_h f_{-+} }{\tilde{w}' +\alpha w_e  + \beta w_h } \:, \nonumber\\
f_{++} &=& \frac{G + \alpha w_e  f_{-+} + \beta w_h  f_{+-}}{\tilde{w}' + w_e + w_h }\:,
\end{eqnarray}
where $\tilde{w}' = w' + D^2 w$. Substituting these expressions into
the two last equations (\ref{anyB}), we arrive at the following
equations for the bright-state populations
\begin{eqnarray}
&&\left( w_{+-} + \beta W \right) f_{+-} - \alpha W  f_{-+} = {\tilde G}_{+-}\:,\nonumber\\
&&- \beta W  f_{+-} +  \left( w_{-+} + \alpha W \right) f_{-+} ={\tilde G}_{-+}\:.\label{f1c}
\end{eqnarray}
Here the effective generation rates
\begin{eqnarray}
&&{\tilde G}_{+-} = G\left( 1 + \frac{ \alpha w_e  }{\tilde{w}' +\alpha w_e  + \beta w_h } + \frac{w_h }{\tilde{w}' + w_e + w_h }\right), \nonumber \\
&&{\tilde G}_{-+} = G\left( 1 + \frac{w_e}{\tilde{w}' + w_e + w_h } + \frac{\beta w_h }{\tilde{w}'+\alpha w_e  + \beta w_h } \right),\nonumber.
\end{eqnarray}
The rates $w_{-+},w_{+-},W$ are defined by
\begin{eqnarray} \label{f1c}
&&w_{-+} = \tilde{w} +\tilde{w}' \left( 1 + \frac{\alpha w_e}{\tilde{w}' + w_e + w_h } + \frac{w_h }{\tilde{w}' + \alpha w_e  + \beta w_h  }\right),  \nonumber\\
&&w_{+-} = \tilde{w} + w' \left( 1 + \frac{w_e}{\tilde{w}' +\alpha  w_e  + \beta w_h } + \frac{\beta w_h }{\tilde{w}' + w_e + w_h }\right),  \nonumber\\
&&W = w_e w_h \left(\frac{1}{\tilde{w}' + \alpha w_e  + \beta w_h } + \frac{1 }{\tilde{w}' + w_e + w_h }\right)\nonumber \:,
\end{eqnarray}
and $\tilde{w} = (C^2 - D^2) w + w'$. As a result we obtain for the bright state populations
\begin{eqnarray}
f_{-+} &=& \frac{ \left(w_{+-} + \beta W \right) {\tilde G}_{-+} + \beta W {\tilde G}_{+-}}{w_{-+}w_{+-} + \left( w_{+-}\alpha  +  w_{+-} \beta \right) W} \:,\nonumber\\
f_{+-} &=& \frac{ \left(w_{-+} + \alpha W \right) {\tilde G}_{+-} + \alpha W {\tilde G}_{-+} }{ w_{-+}w_{+-} + \left( w_{+-} \alpha  +  w_{+-} \beta \right) W}  \:.\label{sol:F}
\end{eqnarray}
The populations of the dark states $--/++$, namely,
$f_{--}$ and $f_{++}$, are expressed via the bright ones $f_{+-}$ and
$f_{-+}$ by virtue of Eq.~\eqref{f2}.

\subsection{Faraday geometry, important limiting case}
\label{subsec:F}

In the longitudinal magnetic field $\mathbf B
\parallel z$ ($C^2=1, D^2=0$), Eq.~(\ref{Pc0}) for the circular polarization degree reduces to
\begin{equation}
\label{Pc:0}
P_c = \frac{f_{-+}-f_{+-}}{f_{-+}+f_{+-}}.
\end{equation}
In this subsection we consider the important limiting case of a sufficiently strong field, so that
\begin{equation}
\label{B:strong}
\alpha,\beta \ll 1.
\end{equation}
Further, we consider a weak nonradiative decay and an efficient spin relaxation from the higher to the lower Zeeman sublevels
\begin{equation}
\label{sr:fast}
w' \ll w \ll w_e,w_h ,
\end{equation}
while the relations between $w',\alpha w_e $ and $\beta w_h $ remain unrestricted. Then the following simplifications are possible
\begin{eqnarray}
&&{\tilde G}_{-+} \approx G \left( 1 + \frac{ \alpha w_e  }{w' + \alpha w_e  + \beta w_h } + \frac{w_h }{w_e + w_h }\right)\:,\nonumber \\
&&{\tilde G}_{+-}\approx G \left( 1 + \frac{ w_e }{w_e  + w_h } + \frac{\beta w_h }{w' +  \alpha w_e  +\beta w_h }\right)\:, \nonumber \\
&&w_{-+} \approx \frac{w'w_e }{w' + \alpha w_e  + \beta w_h }\:, \nonumber \\
&&w_{+-} \approx \frac{w' w_h}{w' + \alpha w_e  + \beta w_h  }\:, \nonumber \\
&& W \approx \frac{w_e w_h}{w' +\alpha  w_e  + \beta w_h }\:,\label{GN}
\end{eqnarray}
with the result
\begin{eqnarray} \label{mod}
&&f_{-+} = \frac{G}{w_h} \left( 1 + \frac{w_e }{w_e + w_h } + 4 \beta \frac{w_h}{w'} \right)\:, \nonumber \\
&&f_{+-} = \frac{G}{w_e} \left( 1 + \frac{w_h }{w_e + w_h } + 4 \alpha\frac{w_e}{w'}\right)\:.\label{app:limit}
\end{eqnarray}

\begin{figure}[hbt]
\includegraphics* [width=8 cm]{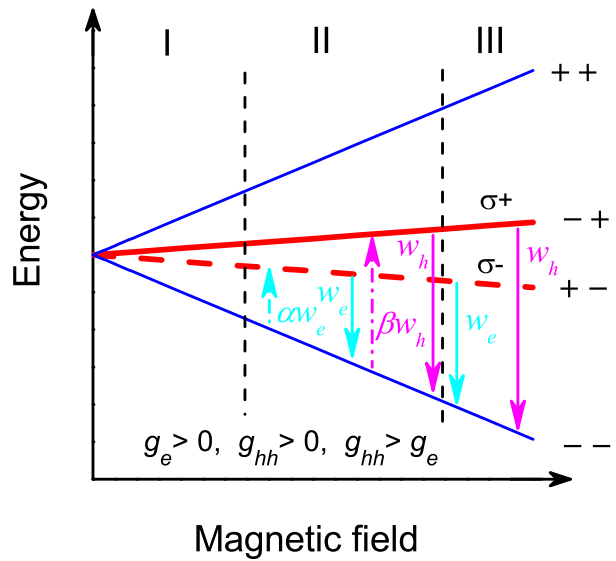}
\caption{\label{fig8} (Color online) Schematic illustration of the
indirect-exciton spin levels in Faraday geometry, $\bm B \parallel z$. The blue lines show optically dark (spin-forbidden) states, the red lines
show spin allowed bright states in $\sigma^+$ (solid
line) and $\sigma^-$ (dash line) polarization. Arrows demonstrate
electron (hole) spin-flip processes increasing and decreasing the energy, respectively,
with rates $\alpha w_e$ ($\beta w_h $) and $w_e$ ($w_h$). The
regions I, II, and III indicate the cases of weak,
moderate, and high magnetic fields, respectively. For definiteness
the case of positive $g_e$, $g_{hh}$ with $g_{hh}>g_e$, which
provides an interpretation of the main experimental findings, is
shown.}
\end{figure}

Equations (\ref{app:limit}) can be obtained by using
simple considerations, namely by (i) analyzing the case of very
strong magnetic fields where only transitions with phonon
emission are possible and (ii) taking into account the phonon
absorption in first-order perturbation theory. First, we
completely neglect transitions accompanied with phonon
absorption, i.e., we set $\alpha, \beta=0$ and obtain from
Eqs.~\eqref{anyB}
\begin{subequations}
\label{zero:order}
\begin{eqnarray}
f_{++} &=& \frac{G}{w_e+w_h}, \: \: \: f_{+-} = \frac{G}{w_e}\left(1 + \frac{w_h}{w_e+w_h}\right)\\
f_{-+} &=& \frac{G}{w_h}\left(1 + \frac{w_e}{w_e+w_h}\right), \: \: \:  f_{--} = 4G/w'.
\end{eqnarray}
\end{subequations}
This result is valid in very strong magnetic fields, shown in Fig.~\ref{fig8} as region III, where $\alpha w_e
\tau_{nr}, \beta w_h \tau_{nr} \ll 1$. Note that since the
polarization is controlled by the transitions from the states
$|-$$+\rangle$ and $|+$$-\rangle$, it shows a dynamic behavior. Indeed,
the sign of polarization is determined by the state which is emptied
slower. If $w_h<w_e$ so that the electron spin flip is fastest, the
population of the state $|+$$-\rangle$ decays faster towards the
ground sublevel $|-$$-\rangle$ and the value of $P_c$ is positive. By
contrast, for $w_h>w_e$, the state $|-$$+\rangle$ decays faster and
$P_c$ becomes negative. This is in agreement with Eq.~\eqref{Pc:0} which
can be recast as
\begin{equation}
\label{Pc:0:1}
P_c =  \frac{1-\zeta^2}{1 + \zeta +\zeta^2}, \quad \zeta = \frac{w_h}{w_e}.
\end{equation}

For moderate magnetic fields according to Eq.~\eqref{B:strong},
shown in Fig.~\ref{fig8} as region II, it is sufficient to
take into account additionally the transitions from the ground
sublevel $| -$$- \rangle$ to the $|-$$+\rangle$, $|+$$-\rangle$
sublevels only. This results in the modification of the bright-state
occupations from \eqref{zero:order} to (\ref{mod}) and of the
polarization from Eq. (\ref{Pc:0}) to
\begin{equation}
\label{Pc:moderate}
P_c =  \frac{w_e^2-w_h^2 + 2\tau_{nr} (\beta - \alpha)(w_e+w_h) w_e w_h}{w_e^2+w_h^2+w_e w_h [ 1+  2 \tau_{nr}(\beta + \alpha) (w_e+w_h)]},
\end{equation}
while $f_{++}, f_{--}$ remain unchanged.

Finally, in the weak-field limit (region I in Fig.~\ref{fig8}) $\alpha w_e \tau_{nr},\beta w_h
\tau_{nr} \gg 1$ the spin relaxation is fast as compared
with the recombination processes, so that the states $|-$$+\rangle$,
$|+$$-\rangle$ become thermally populated and one obtains
\begin{equation}
\label{Pc:therm} P_c = \frac{1-\chi}{1+\chi}, \quad \chi =
\exp{\left(-\frac{(g_e - g_{hh})\mu_B B}{k_{B} T} \right)}.
\end{equation}

In fact, for the parameters of the experimentally studied structure
(see Sec.~V A for the specific values) Eq.~\eqref{Pc:therm} holds
for magnetic fields from $B_z=0$ to $B_z\approx 3$~T at the lowest
accessible temperatures.

\subsection{The Voigt geometry}
\label{subsec:V}

It is also instructive to address the Voigt geometry, $\bm B
\parallel x$, in which case the coefficients $C=D=1/\sqrt{2}$. The
electron spin states then read
\[
| \pm 1/2 \rangle_{\bm B} = \frac{1}{\sqrt{2}} (|1/2\rangle_z \pm |-1/2\rangle_z)\:.
\]
Here the radiative decay rates of all four states are the same.
Moreover,  the exciton states $sj_z$ form two degenerate pairs
($++$, $--$) and ($+-$, $-+$) because the effect of the in-plane
field is disregarded. As a result, according to Eqs.~(\ref{f2}) and
(\ref{sol:F}) the populations $f_{ij}$ are given by

\begin{eqnarray}
\label{sol:V}
&&f_{++} = f_{--} = \frac{G}{\frac{w}{2} + w'} \frac{\frac{w}{2} + w' +2 w_e}{\frac{w}{2} + w' + (1 + \alpha) w_e }\:,\nonumber \\
&&f_{+-} = f_{-+} = \frac{G}{\frac{w}{2} + w'} \frac{\frac{w}{2} + w' +2 \alpha w_e }{\frac{w}{2} + w' + (1 + \alpha) w_e }.
\end{eqnarray}

In the Voigt geometry there is no circular polarization $P_c$ of
the emission. The emission intensity is weakly affected by
the magnetic field.

\subsection{Possible deviation of selection rules}
\label{subsec:selection}

The above model already accounts for the majority of experimental
observations in the present work and in Ref.~\cite{Shamirzaev3}.
However, in order to describe particular features of the observed dynamics
of the exciton luminescence polarization, we need to allow for minor deviations of the selection rules from those
in Eq.~\eqref{1}. As we demonstrate below these deviations are
important mostly in the Faraday geometry. Therefore we
focus on this specific geometry, $\bm B\parallel z$, and specify the
states by the exciton angular momentum component $m_z= s_z + j_z$.
We assume the structure under study to have C$_{2v}$ point
symmetry. In this case the bright exciton states, $\pm 1$, are mixed
with each other, while the dark states, $\pm 2$, can be mixed with
the excited (light-hole exciton) states with $m_z = 0$. As a result, the selection rules
deviate from those described above.

In the simplest model of symmetry reduction from the
$D_{2d}$ point group that is relevant for [001]-grown quantum wells with
symmetric heteropotential and equivalent interfaces, let us take
into account mixing of the heavy-hole, $|\Gamma_8,\pm
3/2\rangle$, and the light-hole, $|\Gamma_8,\mp 1/2\rangle$, states.
Microscopically, the mixing is related to the anisotropy of the
chemical bonds at the interfaces, strain and/or anisotropic
localization~\cite{ivchenko05a} and described by
\begin{equation}
\label{mixing}
\left|{ \pm \frac{3}{2}}\right\rangle = \mathcal N \left(\varphi_{h}(z)\left|\Gamma_8, \pm \frac{3}{2} \right\rangle + \eta  \varphi_{l}(z)\left|\Gamma_8, \mp \frac{1}{2} \right\rangle\right).
\end{equation}
Here  $\varphi_{h}(z)$, $\varphi_{l}(z)$ are the envelope functions
of the heavy- and light-hole states, $\eta$ is the admixture
coefficient, and $\mathcal N = 1/{\sqrt{1+|\eta|^2}}$ is the
normalization constant. The phase of the admixture coefficient is
determined by the choice of axes in the quantum well plane. The
admixture of light hole states results in a modification of
the selection rules: Each of the  ``bright'' states with $|m_z|=1$
is active, both in $\sigma^+$ and $\sigma^-$ polarization, with a strength given by the
ratio of the squared moduli of the matrix elements
\begin{equation}
\label{ratio}
\frac{|M_{\sigma^-}(-1/2,+3/2)|}{|M_{\sigma^+}(-1/2,+3/2)|} = \frac{|M_{\sigma^+}(+1/2,-3/2)|}{|M_{\sigma^-}(+1/2,-3/2)|}= \frac{|\eta|}{\sqrt{3}}.
\end{equation}
The  dark states should become weakly ($|M|^2 \propto |\eta|^2$) optically
active in the $z$ polarization (normal to the QW plane), this
emission is, however, not detected in the experimental geometry. Hence, the
heavy-light hole mixing results in a partial depolarization of the
exciton emission.

Moreover,  besides the mixing of heavy- and light hole described by
Eq.~\eqref{mixing}, we allow for spin-flip processes in
the phonon-assisted recombination which result in an
activation of the dark states, $|m_z|=2$, in the in-plane
polarization. For better agreement with experiment we assume that
the state with $m_z=-2$ is weakly active in the $\sigma^+$
polarization and, reciprocally, the state with $m_z=+2$ is weakly
active in the $\sigma^-$ polarization. As a result, in the Faraday
geometry, we write the degree of exciton luminescence circular
polarization as follows
\begin{equation}
\label{Pc:def}
P_c = \xi \frac{f_{+1} - f_{-1} + C_d (f_{-2} - f_{+2})}{f_{+1} + f_{-1} + {C'_d} (f_{-2} + f_{+2})}.
\end{equation}
Here $\xi = (3 - |\eta|^{2})/(3 + |\eta|^{2})$
is the depolarization factor and the positive coefficients $C_d,
C'_d \ll 1$ account for the emission of the dark states. Setting
$\xi=1$ and $C_d, C'_d = 0$ we return to the strict selection rules
of Eq.~\eqref{1}. Generally, the parameters $\xi$ and $C_d, C'_d$
can depend on the phonon involved in the replica
formation. For example, due to interface effects,  the coupling of
excitons to the short-wavelength phonons involved in the indirect
photoemission can depend on the localization site (e.g., through the
localization energy and, therefore, the localization length) and
this dependence can be different for the TA and LO phonons. Then the
emitting states participating in the TA- and LO-assisted
photoluminescence processes can differ in the heavy-light hole
mixing strength, the depolarization factor $\xi$ and the dark-state
activation factors $C_d, C'_d$. In the following for simplicity we
set $C_d = C'_d$.

\section{Modeling the experimental results}
\label{sec:discussion}

We consider  here a Wannier--Mott exciton formed by an electron in
the doubly-degenerate conduction band and a hole in the
doubly-degenerate valence band. For (001)-grown
heterostructures with zinc-blende lattice, two of the exciton quartet states are bright
and the two other states are dark. The bright excitons can be directly
excited by light and can emit light. The dark excitons are optically
inactive, they recombine with a nonradiative decay rate, have a
longer lifetime and can act as a reservoir of excitons. The
electron-hole exchange interaction results in a splitting between
the bright and dark doublets and, due to reduced symmetry of a
particular nanosystem, in an additional splitting of each doublet.
Moreover, the exciton fine structure can be controlled by an
external magnetic field which  produces a three-fold effect: (i) it
modifies and increases the sublevel splitting, (ii) it mixes the bright
and dark states (at tilted fields or in the Voigt geometry), and
(iii) it induces level crossing or anticrossing.

The population of the split sublevels is arranged by the interplay between recombination
processes and spin flips of either an electron or a hole between the sublevels.  Thus, the exciton photoluminescence
intensity and polarization are governed by the following set of
parameters: the exchange constants, the values and signs of the
electron and hole $g$ factors (together with the strength and
orientation of the magnetic field), the radiative ($\tau_r$) and
nonradiative ($\tau_{nr}$)  recombination times, the spin relaxation rates
($w_e, w_h$) describing the spin flip rates for downward
transitions, i.e., from the upper to the lower Zeeman sublevel, and
the temperature which determines the ratio of upward and downward
transitions. The possibility of various relations between the above
parameters for the quartet excitons offers a vast diversity.

One of the  simplest cases is realized in colloidal CdTe
nanocrystals where due to the strong confinement the bright-dark
exchange splitting exceeds by far the Zeeman energy (in the studied
range of magnetic fields) and the thermal energy $k_{B}T$. Hence, in
this material system at liquid helium temperatures the thermally
induced mixing between bright- and dark-exciton states can be
neglected, and only thermalization between the Zeeman levels of the
bright excitons has to be taken into account \cite{rodina2014}.

The quantum confinement of electrons and holes in InAs/(In,Al,Ga)As
self-assembled quantum dots suppresses the most efficient spin
relaxation mechanisms so that the spin flip processes in the exciton
are much slower than the radiative recombination \cite{Belych2015}.
As for the exchange interaction energy, it is small compared with
the electron and hole Zeeman energies even in a moderate magnetic
field of $B = 0.5$~T.

In studies of pseudo-direct  GaAs/AlAs superlattices where the
$\Gamma$-X mixing is responsible for the nanosecond scale of the
exciton radiative time, see e.g. \cite{baranov3,Dzhioev}, the
exchange and Zeeman splittings are small as compared with the thermal energy at
liquid-helium temperature 4.2~K used in the experiments. Therefore,
the Boltzmann factors~\eqref{Boltz} are close to unity, the
radiative recombination rate is comparable with the spin flip rates,
and level anticrossing occurs in fields of $B {\lesssim}
0.25$~T.

In GaSe crystals and GaSe$_{1-x}$Te$_x$ solid solutions, the
exchange interaction splits the exciton quartet into singlet and
triplet by $\sim 2$~meV with the triplet being additionally split by
$\sim 0.05$~meV in two levels with angular momentum
$z$-components $\pm 1$ (bright exciton) and 0 (dark exciton). One of
the bright-exciton levels anticrosses with the dark level in a field of $B
\approx 0.5$~T.  The radiative and nonradiative recombination times are equal to
1.6 and 10~$\mu$s, respectively, and within this time range the
exciton spin relaxation can be disregarded \cite{Starukhin}.

In the GaAs/AlAs heterostuctures studied in this paper and in
Ref.~\cite{Shamirzaev3}, the  quartet of indirect exciton states is
formed by electrons in the conduction X$_{x}$-X$_y$ valleys and
$\Gamma$-point holes. As compared to the previously analyzed
nanosystems, the distinctive features of such the excitons are:
{(i)}~the radiative exciton times aount to $\sim$~ms, much longer
than the spin relaxation times:
$$\tau_{se} = w_e^{-1}, \tau_{sh} = w_h^{-1} \ll \tau_r,
\tau_{nr}\:,$$ {(ii)}~application of magnetic fields can significantly reduce the rates of the
electron and hole upward spin flips at low temperatures and make them comparable with
the nonradiative decay rate. The level anticrossing in weak
magnetic fields is out of the scope of the present work (it takes place
at $B<100$~mT). This particular relation of parameters yields a
pronounced interplay of the bright and dark exciton states in the
photoluminescence and to the non-trivial circular polarization
dependence of the PL on magnetic field and temperature. Below, in
Sec.~\ref{subsec:modelling}, we outline the restrictions on the model
parameters needed to successfully reproduce the experimental
findings. The theoretical estimate of the heavy-hole $g$-factor is
given in Sec.~\ref{subsec:ghh}. Further, Sec.~\ref{sec:variants}
outlines possible situations which can be realized depending on the
relative signs and magnitudes of the electron and hole Land\'{e}
factors, where one can expect a strong dependence of $P_c$ on
magnetic field and temperature.

\subsection{Parameters used for fitting}
\label{subsec:modelling}

The exciton recombination in the studied GaAs/AlAs QWs described by
Eqs.~(\ref{anyB}) is governed by eight parameters: the electron
($g_e$) and heavy hole ($g_{hh}$) $g$ factors, the recombination
times of bright ($\tau_{r} = 1/w$) and dark ($\tau_{nr} =
1/w^{\prime}$) excitons, the spin relaxation times for electrons
($\tau_{se}$) and heavy holes ($\tau_{sh}$), the depolarization
factor $\xi$, and the dark-state activation factor $C_d$. Some of these
parameters  can be directly measured in experiment, others can be evaluated
from fits of various experimental dependencies, or at least
the ratio of parameters and their possible ranges of values can be found.

Let us start from well defined parameters. Due to the large band gap
at the $X$ point, the spin-orbit contribution to the electron $g$
factor is vanishingly small~\cite{ivchenko05a,Yugova}. As a result,
the electron $g$ factor, $g_{e}$, is isotropic and its value almost
coincides with the free-electron Land\'e factor of
$+2.0$~\cite{Debus}. The recombination times of the bright and dark
excitons were unambiguously determined in Ref.~\cite{Shamirzaev3},
$\tau_{r}=0.34$~ms and $\tau_{nr}=8.5$~ms, by analyzing the PL
intensity dynamics and its variation with magnetic field. The
solution of Eqs.~\eqref{anyB} for pulsed excitation was used to
calculate the $P_c(t)$-dynamics with Eq.~\eqref{Pc0}. The analysis
of the calculated results demonstrates that the polarization degree
rise-time is uniquely determined by the shortest among the electron
and hole spin relaxation times and does not depend on other
parameters. It will be shown in Sec.~VI~B, that the sign of $P_c$ in
the initial growth stage is unambiguously controlled by the
$\tau_{se}/\tau_{sh}$ ratio. Therefore, the fit of the $P_c(t)$
dynamics presented in Fig.~\ref{fig7}(b) allows us to conclude that
$\tau_{se}$ is longer than $\tau_{sh}=3\pm 0.5$~$\mu$s, which is
valid for all magnetic field strengths and orientations.

Thus, we have four variable parameters to describe the experimental
findings: the heavy hole longitudinal $g$ factor $g_{hh}$, the
electron spin relaxation time $\tau_{se}$, and the factors $\xi$ and
$C_d$. However, the possibilities for variation of these parameters
are very limited. The factor $\xi$ is unambiguously determined by
the saturation level of $P_c$ in tilted magnetic field as shown in
Figs.~\ref{fig5} and ~\ref{fig6}. It equals to 0.75 for the
LO$_{\mathrm {AlAs}}$ phonon assisted line. The factor $C_d$
together with $g_{hh}$ are responsible for the magnetic field value
where the maximum value of the circular polarization degree ,
$|P_{c,max}|$, is reached in Faraday geometry, see
Figs.~\ref{fig2},~\ref{fig3}, and ~\ref{fig5}. An additional
restriction is imposed on the heavy hole longitudinal $g_{hh}$
factor. It cannot be smaller than $+2.5$ as recently established
from the PL intensity dependence on magnetic
field~\cite{Shamirzaev3}. Finally the electron spin relaxation time
$\tau_{se}$ defines the slope of the $P_c(B)$ decrease in strong
magnetic fields applied in the Faraday geometry as shown in
Figs.~\ref{fig2},~\ref{fig3}, and ~\ref{fig5}.

The best fits of $P_c(B)$ and the PL decay at different temperatures,
magnetic fields and orientations are shown by the lines in
Figs.~\ref{fig2}, \ref{fig3}, \ref{fig4}, \ref{fig5}, and
\ref{fig6}. All of them are obtained with the following set of parameters:
$\tau_{se}=33\pm 1$~$\mu$s, $C_d=0.001$, and $g_{hh} = +3.5\pm 0.1$.
In Sec.~\ref{subsec:ghh} we show that the large positive value of
$g_{hh}$ is consistent with a simple estimation based on an
effective Hamiltonian approach and with experimental data for
short-period GaAs/AlAs superlattices. The parameters used for the
model calculations are collected in Table~\ref{tab:parameters}. It is
remarkable that almost perfect quantitative description of
all measured experimental dependencies is achieved, which confirms
the validity of the used model.

\begin{table}[h]
    \caption{Parameters for the studied ultrathin GaAs/AlAs QWs evaluated either experimentally or from the best fit to the experimental data.}
  \begin{ruledtabular}
        \begin{tabular}{lcc}
            Parameter& Value & Comment \\ \hline
            $g_{e}$ & $+2.0$ & \cite{Debus} \\
            $g_{hh}$ & $+3.5\pm 0.1$ & best fit\\
            $\tau_{r}$ & 0.34~ms & \cite{Shamirzaev3}\\
            $\tau_{nr}$ & 8.5~ms & \cite{Shamirzaev3} \\
            $\tau_{sh}$ & $3\pm 0.5$~$\mu$s & fit in Fig.~\ref{fig7}(b)\\
            $\tau_{se}$ & $33\pm 1$~$\mu$s & best fit \\
            $\xi$ & 0.75 & Figs.~\ref{fig5} and ~\ref{fig6}  \\
            $C_d$ & 0.001 & best fit \\
 \end{tabular}
  \end{ruledtabular}
    \label{tab:parameters}
\end{table}

Fitting the experimental data shows that in the studied GaAs/AlAs
QW at $T=1.8$~K the regime of weak magnetic fields is valid for $0 <
B < 3$~T and that of moderate magnetic fields for 4~T$ < B < $~10~T, see
also Fig.~\ref{fig8} and Sec.~\ref{subsec:F}. One can see in
Figs.~\ref{fig3} that the regime of strong magnetic fields, where we can
set $\alpha, \beta=0$, is reached for $B > 12$~T.

Using $C_d=0.001$ and depolarization factors $\xi$ equal to 0.75
and 0.6, respectively, we obtain excellent fits for all
dependencies of the LO$_{\mathrm {AlAs}}$ and TA$_{\mathrm {AlAs}}$
phonon assisted lines. In fact, the $P_c(B)$ dependencies of the LO
and TA phonon replicas in Fig.~\ref{fig2} perfectly match each other
after scaling by $1/\xi$, which validates the introduction of the
replica-dependent depolarization factor in our model. We note that a
reasonable description of the no-phonon line is possible for
$\xi=0.31$ and $C_d =0.001$, but the fit accuracy is
significantly smaller for the NP line as compared with the phonon replicas. We
exclude this line from consideration because the origin of the NP
luminescence is not fully established in our
sample~\cite{Shamirzaev3}.

Note that in the fit of the experimental data presented above we
assumed that the generation rates in all states are equal $G_{sj_z}
\equiv G$. We have checked that accounting for the thermalization of
electrons and holes between their spin states in the course of
relaxation does not allow us to reproduce the dynamics of
polarization and, moreover, in sufficiently strong magnetic fields
makes the contribution of the dark states too strong to account for
the observed values.

\subsection{Theoretical estimate of the heavy-hole $g$ factor}
\label{subsec:ghh}

To calculate the exciton states and their $g$ factor in a
monolayer-thick GaAs/AlAs QW one needs an atomistic computation
based on a microscopic theory such as a tight-binding model. This
is beyond the scope of the present work. Instead, in order to get a
crude estimate of the heavy-hole $g$ factor in the ultrathin QW we
apply an effective Hamiltonian and take into account the
magnetic field-induced mixing of the heavy- and light-hole
states. The monolayer-thick QW is modeled by a rectangular
heteropotential $V(z)$ with well width $a_0=0.565$~nm and
GaAs/AlAs valence band offset of $\Delta
V=0.53$~eV~\cite{Vurgaftman02}: $V(z) = 0$ inside the well and $V(z)
= \Delta V$ in the barriers. In such a well there are only one
heavy-hole ($hh$1) and one light-hole ($lh$1) quantum-confined subbands. The magnetic-field-induced mixing calculated in the framework of the
Luttinger Hamiltonian takes place between subbands of different
parity. Therefore, in the evaluation of $g_{hh}$ we have to take into
account the linear-in-${\bm k}$ mixing of the $hh$1 subband states
with the continuum of the light-hole states, where ${\bm k}$ is the
in-plane hole wave vector. Following Ref.~\cite{Durnev2014} (see also Refs.~\cite{efros:g,Durnev2012}) we
represent the $g$ factor of the heavy hole as
\begin{equation}
\label{ghh} g_{hh} =-6\tilde{\kappa} + 4\sqrt{3} a_0 \int dz S_h(z)
\left\{ \gamma_3 \frac{d}{dz} \right\}_s C_h(z),
\end{equation}
where $\tilde{\kappa}$ is the average magnetic Luttinger parameter
between the two materials, $\{ab\}_s = (ab+ba)/2$ is the symmetrized
product of operators, $C_h(z)$ is the envelope wavefunction in the
ground $hh$1 subband (with the quantum-confined energy $E_{hh1}$) and the function $S_h(z)$ describing the $\bm
k$-linear light-hole admixture is defined by the equation
\begin{multline}
\label{Sh}
\left[-\frac{\hbar^2}{2m_0} \frac{d}{dz} (\gamma_1+2\gamma_2) \frac{d}{dz} + V(z) - E_{hh1}\right] S_h(z) =\\
 -\frac{\sqrt{3}\hbar^2}{m_0 a_0} \left\{ \gamma_3 \frac{d}{dz}\right\}_s C_h(z),
\end{multline}
with $\gamma_i$ ($i=1,2,3$) being the  Luttinger band parameters.
Making use of this shallow quantum-well model we recast $g_{hh}$ in
the form
\begin{equation}
\label{ghh:3}
g_{hh} = -6\kappa + \frac{\gamma_3^2}{\gamma_1- 2\gamma_2} \frac{\nu^2}{(1+\nu)^2} \approx 2.67,
\end{equation}
where $\nu = \sqrt{(\gamma_1 - 2\gamma_2)/(\gamma_1 + 2\gamma_2)}$
and all the Luttinger parameters correspond to those of the AlAs material:
$\gamma_1 = 3.76$, $\gamma_2 = 0.82$,  $\gamma_3=1.42$, and
$\kappa=0.12$~\cite{Vurgaftman02}. It is worth stressing, that the
final value of $g_{hh}$ differs not only in magnitude but also in
sign from the averaged bulk value of $- 6 \kappa$. The positive sign
of $g_{hh}$ is determined by the dominant contribution to the $g$
factor arising due to the ${\bm k}$-linear admixture of continuum
light-hole states to the confined heavy-hole states. Thus, $g_{hh}$
is positive and exceeds the value of $g_e \approx 2$ expected for
electrons in the $X$-valleys of AlAs. This result
qualitatively agrees with effective Hamiltonian
calculations \cite{kiselev:g} and results of experimental
magneto-optical studies of GaAs/AlAs short-period superlattices
\cite{VanKesteren,baranov,baranov1,Khitrova,baranov2}. Particularly,
Baranov et al. \cite{baranov} obtained $g_{e} = 1.90$, $g_{hh} =
3.2$ in a 0.74~nm/1.43~nm superlattice, and $g_{e} = 1.88$,
$g_{hh} = 3.14$ in a 1.05~nm/2.99~nm sample.

\section{Model predictions for structures with different sets of parameters}
\label{sec:variants}

In this section, using the developed model we simulate the
experimental appearances in structures with different sets of
parameters in order to highlight the variability of possible
experimental realizations. This can serve as a guide for the further search for
systems with tailored dependencies of the polarization degree on magnetic field and temperature.
We consider scenarios where the
electron and hole spin relaxation times are shorter than the
recombination time of the bright exciton, to avoid complications
related with transient spin relaxation processes. Hereinafter we
choose for definiteness the following values of the exciton
recombination times: $\tau_{r}=0.5$~ms and $\tau_{nr}=10$~ms in all calculations. For simplicity, we set the depolarization
factor $\xi=1$. As before we neglect the
electron-hole exchange interaction.

The PL intensity and exciton dynamics depend strongly on the signs and
magnitudes of the electron and heavy-hole $g$
factors~\cite{Shamirzaev3}. Let us introduce, following the
approach used in Ref.~\onlinecite{bartsch11}, the effective $g$
factors of the bright and dark excitons in the Faraday geometry as
\[
g_{b} = g_{hh} - g_{e}, \quad g_{d} = g_{hh} + g_e,
\]
respectively. $g_b$ describes the splitting of the excitons active
in $\sigma^+$ and $\sigma^-$ polarization as $E_{-+} - E_{+-} = g_b
\mu_B B_z$, while $g_{d}$ gives the splitting of the states $++$ and
$--$. Provided that $|g_{b}| > |g_{d}|$ (\emph{case} 1, illustrated
in Fig.~\ref{fig8}) the Zeeman  splitting of the bright excitons
exceeds the splitting of the dark states, so that in a magnetic
field $B_z\ne 0$ the lowest state is a bright exciton. By contrast,
if $|g_b|<|g_d|$ (\emph{case} 2, Fig.~\ref{fig9}), the exciton state
lowest in energy is dark.

In the studied GaAs/AlAs QW both $g_e$ and $g_{hh}$ are
positive, hence, $g_b<g_d$ and the case 2 is realized. Below we
analyze various situations for different relations
between the kinetic parameters $\tau_{se}$ and $\tau_{sh}$.
For illustration of the model predictions we select
those dependencies that clearly demonstrate distinct features for the different chosen sets
of parameters: the dynamics of the magnetic-field-induced circular polarization
degree, the dependence of $P_{c}$ on temperature and magnetic field orientation
at fixed high field strength, and $P_{c}(B)$ in longitudinal
magnetic field and in tilted field geometry with $\theta= 45^\circ$.

\subsection{Case 1, $|g_b|>|g_d|$}
\label{subsec:case1}

\begin{figure}[]
\includegraphics* [width=\linewidth]{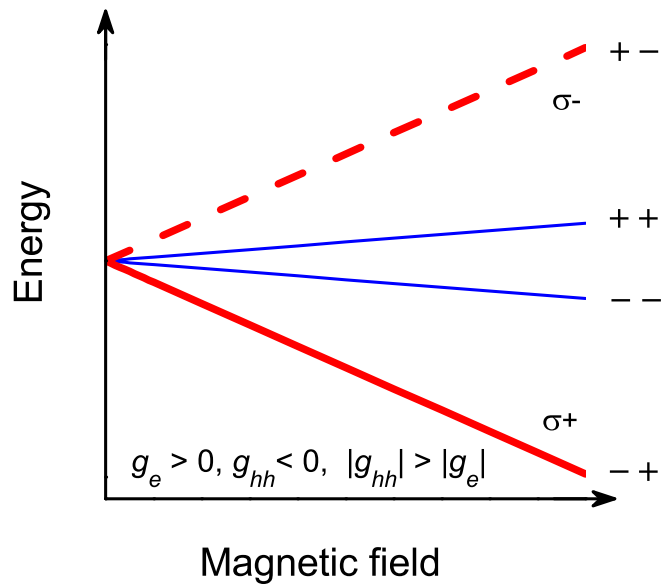}
\caption{\label{fig9} (Color online) Schematics of the exciton spin
levels in the Faraday geometry $\bm B
\parallel z$ for the case 1: $g_e>0$, $g_{hh}<0$, and
$|g_{hh}|>|g_e|$. Blue lines show optically dark (spin-forbidden)
states, red lines show spin allowed bright states active in
$\sigma^+$ (solid line) and $\sigma^-$ (dash line) polarization.}
\end{figure}

\begin{figure}[]
\includegraphics[width=\linewidth]{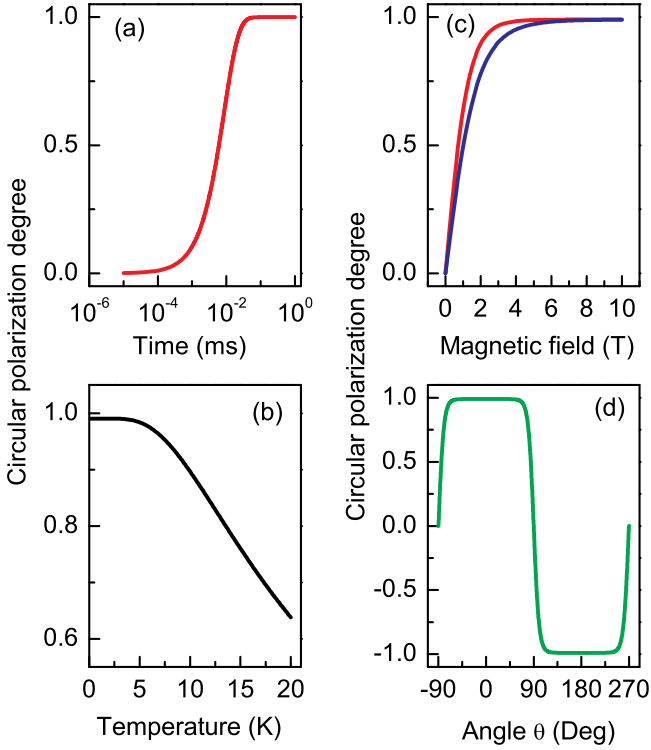}
\caption{\label{fig10} (Color online) Calculated dependencies for
$P_{c}(B)$: (a) Dynamics of ${P_c}$ measured in the Faraday
geometry. (b) ${P_c}$ as function of temperature. (c) ${P_c}$ as
function of longitudinal (red line) and tilted by $45^\circ$ (blue
line) magnetic field. (d) ${P_c}$ as function of tilt angle.
Parameters  used in the calculations: $T=2$~K, $B=10$~T, $g_e=+2$,
$g_{hh}=-3$, $C_d=0.001$, $\tau_{se}=50~\mu$s, $\tau_{sh}=5~\mu$s,
$\tau_{r}=0.5$~ms, and $\tau_{nr}=10$~ms.}
\end{figure}

For definiteness we assume that $g_e>0$, $g_{hh}<0$, and
$|g_{hh}|>|g_e|$. The only configuration of the exciton
spin levels that is possible in this case in the Faraday geometry is
schematically shown in Fig.~\ref{fig9}. The dependencies of the selected quantities
calculated for the set of parameters given in the figure caption
are shown in Fig.~\ref{fig10}. When we vary the parameters across wide
ranges, namely $g_e$ from 0.1 to 2, $g_{hh}$ from $-0.3$ to $-3$, $C_d$ from zero to 0.01, and
$\tau_{se}$ and $\tau_{sh}$ from $0.01~\mu$s to 0.1~ms for fixed
values of $T$ and $B$, all dependencies demonstrate the same
trends: (i) after the excitation pulse $P_{c}$ increases with time up to
saturation, with the corresponding time constant being the shorter one of the electron and
hole spin relaxation times [Fig.~\ref{fig10}(a)]; (ii) $P_{c}$
monotonically decreases with increasing temperature, in accordance
with a Boltzmann population of the bright exciton levels
[Fig.~\ref{fig10}(b)]; (iii) the angular dependence of $P_{c}$ reflects
the Zeeman splitting of the bright exciton levels by projecting the
magnetic field on the growth axis of the structure
[Figs.~\ref{fig10}(c) and \ref{fig10}(d)]. In fact, in this regime the recombination
and spin dynamics behaviors of the four-level system reduce to those
of a two-level system involving bright states only. The dark states are important as intermediate states
only in the course of electron and hole spin relaxation and
the reservoir of dark excitons is not formed due to efficient downward spin flip processes
towards the optically active exciton state of lowest energy~\cite{Shamirzaev3}, see Fig.~\ref{fig9}.

\subsection{Case 2, $|g_b| < |g_d|$}
\label{subsec:case2}

\begin{figure}[]
\includegraphics* [width=\linewidth]{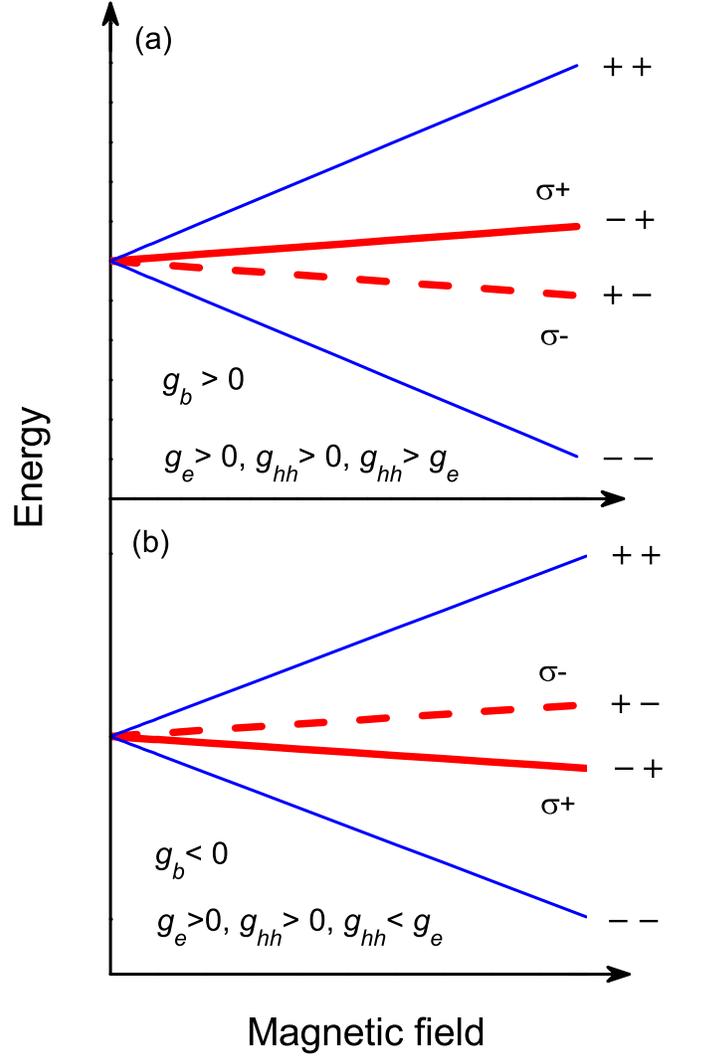}
\caption{\label{fig11} (Color online) Schematics of exciton spin
levels in the Faraday geometry for the case 2: $g_e >0$, $g_{hh}>0$, on one case
(a) $g_{hh} > g_e $ and in the other case (b) $g_{hh} < g_e$. The blue lines show the optically
dark (spin-forbidden) states, the red lines show the spin-allowed bright
states active in $\sigma^+$ (solid line) and $\sigma^-$ (dashed line)
polarizations.}
\end{figure}

For definiteness we set here $g_e>0$, $g_{hh}>0$. The magnetic
field applied in the Faraday geometry quenches the PL then, as the dark
exciton becomes the lowest energy state. Depending on the system parameters
the exciton PL and polarization behaviors become multifaceted. In
contrast to the case 1 two possible configurations of exciton spin
levels, which differ by the sign of the bright exciton $g$ factor, can be
distinguished here, as shown in Fig.~\ref{fig11}.

From Sec.~\ref{subsec:F}, in the range of weak magnetic fields, the
$P_{c}$ behavior is similar to the case of a two-level system and is
determined by the Boltzmann population of the bright exciton levels.
Therefore, the most interesting scenarios can be found in the ranges
of moderate and strong magnetic fields. Since these ranges are
determined by the ratio of magnetic ($g\mu_B B_z$) and thermal
($k_{B}T$) energies we use the following sets of parameters that
provide corresponding conditions: $T=2$~K, $B=10$~T, $g_e=+2$,
$g_{hh}=+2.5$ to provide a positive sign of $g_b$ and $g_{hh}=+1.5$
for a negative sign of $g_b$. Even the strict selection rules given
by Eq.~\eqref{1} with $C_d=0$ lead to an interesting behavior of the
emission circular polarization degree $P_c$ as function of time $t$,
magnetic field $B$ and temperature~$T$.

\begin{figure}[]
\includegraphics* [width=\linewidth]{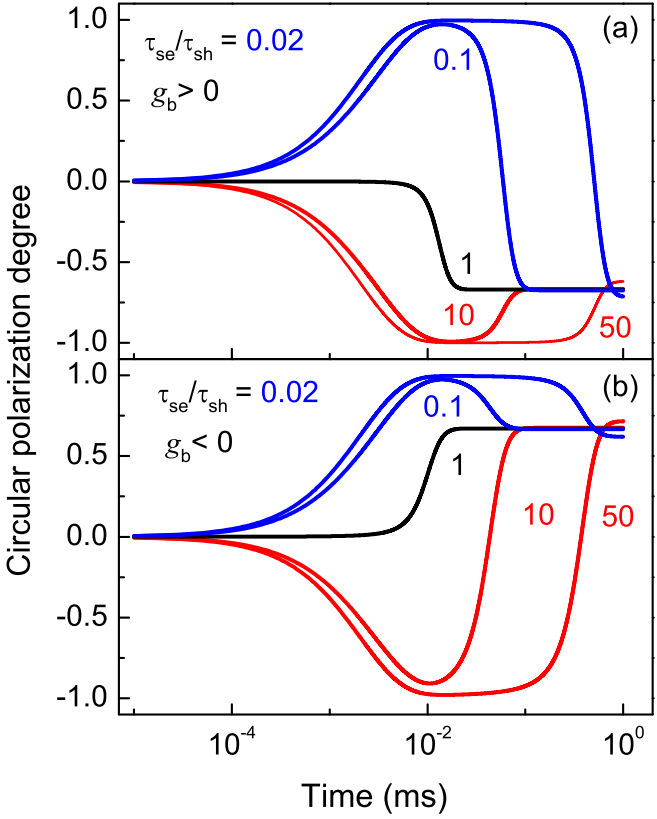}
\caption{\label{fig12} (Color online) Calculated dynamics of
$P_{c}(t)$ for different ratios of $\tau_{se} / \tau_{sh}$ with the following fixed parameter set:
$T=2$~K, $B=10$~T, $C_d=0$, $\tau_{r}=0.5$~ms, $\tau_{nr}=10$~ms,
and $g_e=+2$. (a) $g_{hh}=+2.5$, $g_b>0$ and (b) $g_{hh}=+1.5$,
$g_b<0$. We use in the calculations $\tau_{se} = 1~\mu$s for $\tau_{se}
/ \tau_{sh} \le 1$ and $\tau_{sh} = 1~\mu$s for $\tau_{se} /
\tau_{sh} \ge 1$.}
\end{figure}

Let us start with the $P_{c}(t)$ dynamics calculated for different
ratios of $\tau_{se} / \tau_{sh}$ shown in Fig.~\ref{fig12}. These functions
are nonmonotonic. The rise stage is controlled by the shortest spin relaxation time,
while the decay stage is determined by the longest one of the electron and hole spin
relaxation times. The sign of $P_c(t)$ in the rise stage is
determined by the $\tau_{se} / \tau_{sh}$ ratio and does not depend on the $g_b$ sign.
This is because the bright states $+-$ and $-+$ are initially equally populated and their
population state dynamics is controlled by the kinetic parameters.  Rise-time and decay-time of
the polarization degree are correspondingly determined by the shortest and
longest among the electron and hole spin relaxation times.

\begin{figure}[]
\includegraphics* [width=\linewidth]{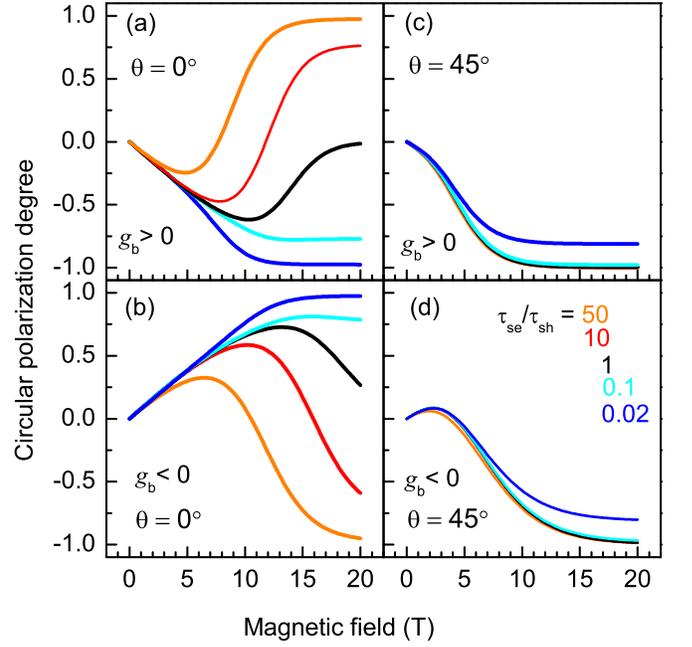}
\caption{\label{fig13} (Color online)  $P_{c}(B)$ dependencies
calculated for different ratios of $\tau_{se} / \tau_{sh}$
[presented in panel (d)] for $T=2$~K, $C_d=0$, $\tau_{r}=0.5$~ms,
$\tau_{nr}=10$~ms, and $g_e=+2$. (a) Faraday geometry: $g_{hh}=+2.5$
and $g_b>0$. (b) Faraday geometry: $g_{hh}=+1.5$ and $g_b<0$. (c)
Tilted geometry: $g_{hh}=+2.5$ and $g_b>0$. (d)
Tilted geometry: $g_{hh}=+1.5$ and $g_b<0$. We
use in the calculations $\tau_{se} = 1~\mu$s for $\tau_{se} / \tau_{sh}
\le 1$ and $\tau_{sh} = 1~\mu$s for $\tau_{se} / \tau_{sh} \ge 1$. Note that as before
only the Zeeman effect caused by the $z$ component of the magnetic field is taken into account for the heavy holes.}
\end{figure}

\begin{figure}[t]
\includegraphics* [width=\linewidth]{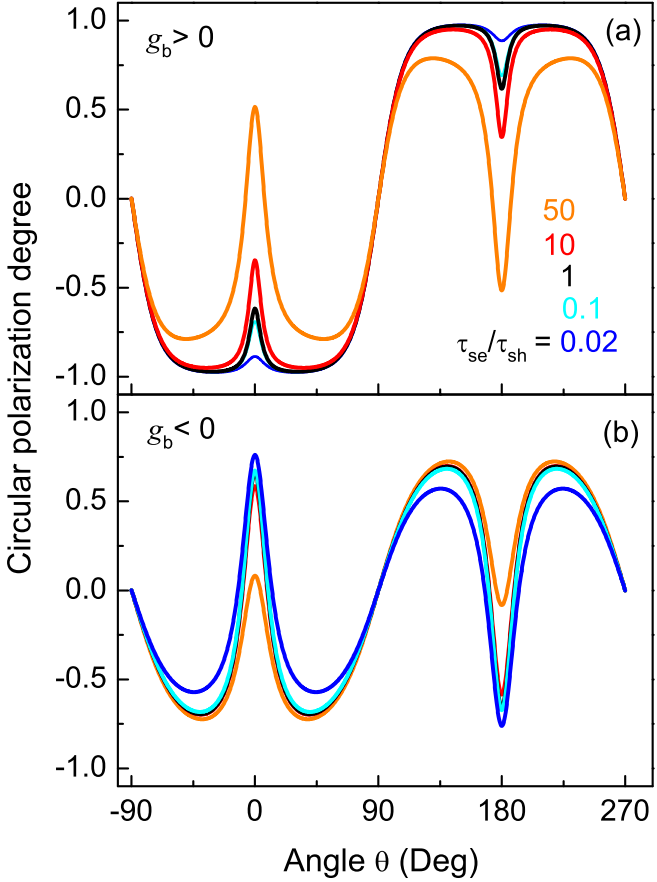}
\caption{\label{fig14} (Color online) Dependencies of
${P_c}(\theta)$ calculated for $T=2$~K and $B=10$~T, $C_d=0$,
$\tau_{r}=0.5$~ms, $\tau_{nr}=10$~ms, and $g_e=+2$ for different
ratios of $\tau_{se} / \tau_{sh}$ [panel (a)]. (a)
$g_{hh}=+2.5$ and a longitudinal value of $g_b>0$
and (b) $g_{hh}=+2.5$ and a longitudinal value of
$g_b<0$. We use in the calculations $\tau_{se} = 1~\mu$s for $\tau_{se}
/ \tau_{sh} \le 1$ and $\tau_{sh} = 1~\mu$s for $\tau_{se} /
\tau_{sh} \ge 1$. Note that as before only the Zeeman effect caused by
the $z$ component of the magnetic field is taken into account for the heavy holes.}
\end{figure}

The circular polarization dependence on magnetic field, $P_{c}(B)$, calculated for the Faraday and for tilted geometries at $T=2$~K are shown in Fig.~\ref{fig13}. Most spectacularly a nonmonotonic behavior with sign reversal is expected for the Faraday
geometry, see Figs.~\ref{fig13}(a) and \ref{fig13}(b). In weak
magnetic fields the slope of $P_{c}(B)$ and its sign are controlled
by $g_b$. One can see that the dependencies calculated for a varying
$\tau_{se} / \tau_{sh}$ ratio closely follow each other in agreement with Eq.~\eqref{Pc:therm}. By contrast, in
strong magnetic fields the $P_{c}$ sign and value are again
determined by the $\tau_{se} / \tau_{sh}$ ratio, see Eqs.~\eqref{Pc:0:1} and \eqref{Pc:moderate}. Depending on this
ratio the $P_{c}(B)$ dependence can be either monotonic reaching
saturation, or nonmonotonic with a sign reversal, but also reaching
saturation in strong fields. In the tilted geometry
($\theta=45^\circ$) the $P_{c}(B)$ dependences demonstrate only a weak maximum
and follow closely each other for the varying values of the $\tau_{se}
/\tau_{sh}$ ratio and the sign of $g_b$, see Figs.~\ref{fig13}(c) and
\ref{fig13}(d). In fact, as shown in Fig.~\ref{fig14},
a significant change of the magnetic field-induced circular polarization
degree occurs for the angles $\theta$ around the Faraday geometry
($\theta \approx 0^\circ$ and $\theta \approx 180^\circ$), while the
mixing of states in tilted geometry neutralizes the contribution
of carrier spin-flips on the polarization degree.

\begin{figure}[]
\includegraphics* [width=\linewidth]{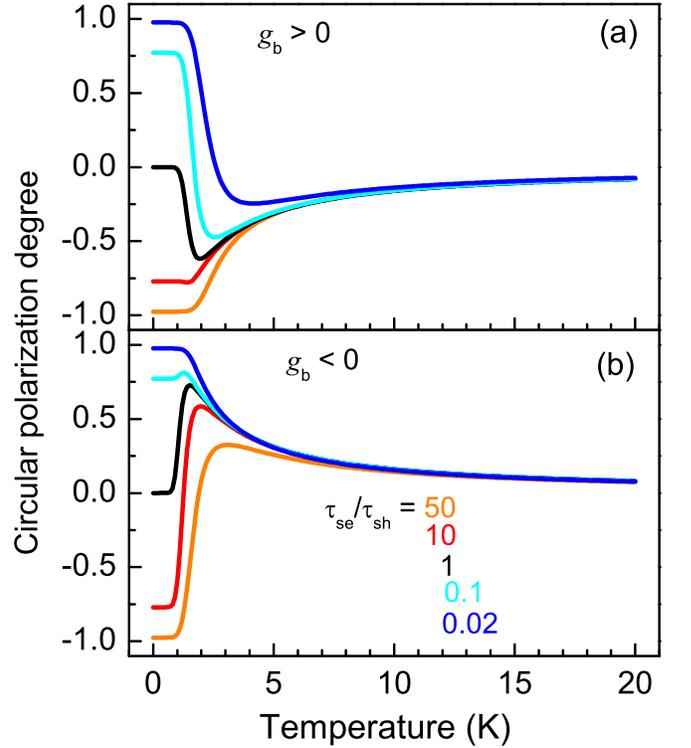} \caption{\label{fig15} (Color online)
Temperature dependencies of ${P_c}$ calculated for different ratios
of $\tau_{se} / \tau_{sh}$ (given in panel (b)) in the Faraday geometry
at $B=10$~T, $C_d=0$, $\tau_{r}=0.5$~ms, $\tau_{nr}=10$~ms, and
$g_e=+2$. (a) $g_{hh}=+2.5$, $g_b>0$ and (b) $g_{hh}=+1.5$, $g_b<0$.
We use in the calculations $\tau_{se} = 1~\mu$s for $\tau_{se} /
\tau_{sh} \le 1$ and $\tau_{sh} = 1~\mu$s for $\tau_{se} / \tau_{sh}
\ge 1$.}
\end{figure}

The temperature dependencies for ${P_c}$ in the Faraday geometry are
also nonmonotonic and depend strongly on the kinetic parameter
$\tau_{se} / \tau_{sh}$ ratio in the range of strong magnetic
fields, see Fig.~\ref{fig15}. At high temperatures (above 7~K) where
the ratio $g\mu_B B_z/k_{B}T \ll 1$ we return to the case of
Boltzmann population of the bright exciton sublevels,
Eq.~\eqref{Pc:therm}.

\subsection{Case 2. Role of deviation from strict selection rules for optical transitions}
\label{subsec:deviation}

\begin{figure}[t]
\includegraphics* [width=\linewidth]{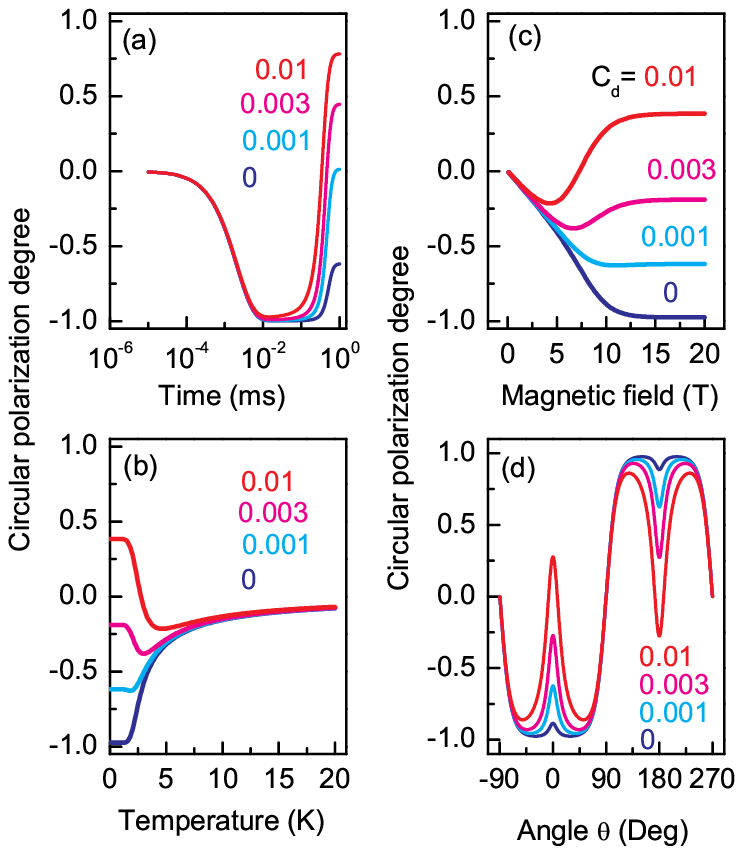}
\caption{\label{fig16} (Color online) Calculated dependencies of
${P_c}$ for different factors $C_d$ of the dark exciton emission
activation. (a) ${P_c}(t)$ dynamics in the Faraday geometry. (b)
${P_c}(T)$. (c) ${P_c}(B)$ in the Faraday geometry. (d)
${P_c}(\theta)$. Used parameters in the calculations: $T=2$~K, $B=10$~T, $g_e=+2$,
$g_{hh}=+2.5$, $\tau_{se}=50~\mu$s, $\tau_{sh}=1~\mu$s,
$\tau_{r}=0.5$~ms, and $\tau_{nr}=10$~ms.  }
\end{figure}

Finally, let us have a closer look into the effect expected from a
small deviation from the optical selection rules. Here interesting
scenarios can arise for moderate and strong magnetic fields when the
polarization of the lower dark state emission related with the coefficient $C_d$ in Eq.~\eqref{Pc:def} differs from that of the lower bright state.
We assume, similar to the situation revealed for
the studied GaAs/AlAs QW, that the state with $m_z=-2$ is weakly
active in the $\sigma^+$ polarization and, vice versa, the state
with $m_z=+2$ is weakly active in the $\sigma^-$ polarization. The
corresponding configuration of the exciton spin levels is shown in
Fig.~\ref{fig11}(a). The magnetic field dependencies of ${P_c}$
calculated at $T=$2~K and for $\tau_{se} / \tau_{sh}$ =50 using different
values of the dark exciton emission activation factor $C_d$ are shown in
Fig.~\ref{fig16}. The most significant changes occur again for the
angles $\theta$ close to the Faraday geometry ($\theta \approx 0^\circ$
and $\theta \approx 180^\circ$) at low temperatures, as
clearly seen in Fig.~\ref{fig16}(b) and \ref{fig16}(d). One can see,
that the factor $C_d$ strongly affects the dynamics of $P_{c}(t)$
in Fig.~\ref{fig16}(a) and the ${P_c} (B)$ dependencies in
Fig.~\ref{fig16}(c).

The model calculation results presented in Sec.\ref{sec:variants} demonstrate a
great  variety of possible experimental dependences of $P_c$
demonstrating that the magnetic-field-induced circular polarization
degree is a sensitive tool to decode the spin structure and the spin
dynamics of excitons in low dimensional structures.

\section{Conclusion}
\label{sec:conclusions}

We investigated experimentally and theoretically the magneto-optical properties of a two-monolayer-thick GaAs/AlAs
quantum well, which is indirect both in real and in $\bm k$ space. The exciton spin dynamics have been addressed through the emission circular polarization induced by external magnetic fields. The extremely long exciton recombination time provided by the spatial separation of electron and hole due to the type-II band alignment in combination with electrons in the X valley results in a novel physical situations. The spin relaxation times of electrons and holes are remarkably long, $\tau_{se}=33~\mu$s and $\tau_{sh}=3~\mu$s, compared to, e.g., type-I GaAs/(Al,Ga)As QWs. The combination of the system parameters with the nonradiative recombination time being long compared with the radiative one and the possibility to control the rate of
transitions from the lower to the upper Zeeman levels by the magnetic fields result in unusual dependences of the emission polarization on
the magnetic field and temperature. A kinetic equation model which accounts for the dynamics of the quartet of
bright and dark exciton states in the QW with C$_{2v}$ point symmetry provided by the heavy-light hole mixing has been developed.
A perfect quantitative description of all experimental data is obtained with just a few variable parameters which can be unambiguously determined from the experimental data. We demonstrate that the magnetic field-induced circular polarization can be controlled either by thermodynamical parameters, i.e., the ratio of the exciton Zeeman splitting and the thermal energy, or by kinetic parameters, i.e., the relations between the various relaxation times in the system, depending on temperature and magnetic field. Furthermore, we have extended the model calculations to varying parameters sets in order to highlight the role of specific parameters in the experimental appearances. The developed approach can readily be used for investigation of the spin dynamics in semiconductor quantum well and quantum dot structures with indirect band gap either in real or in $\bm k$ space, or in both of them.

{\bf Acknowledgements} We thank M.V.~Durnev for valuable discussions. This work was supported by the German
Ministry of Education and Research (BMBF) (FKZ: 05K13PE1), the
Deutsche Forschungsgemeinschaft and the Russian Foundation for Basic
Research via ICRC TRR160, the Russian Foundation for Basic Research
(Grants No. 16-02-00242, 15-52-12012, and 17-02-00383), the Russian Federation Government Grant No. 14.Z50.31.0021
(leading scientist M.~Bayer), and by the Act 211 Government of the
Russian Federation (Contract No. 02.A03.21.0006).


\begin{thebibliography}{}
\bibitem{choiceofGaAs} We chose for this study a very thin GaAs QW in order to be able to compare its properties to another InAs/AlAs model QW structure, which has an indirect band gap, but a type-I band alignment. Due to strain, high quality InAs/AlAs QWs can be grown with a few monolayer thickness only. This does not represent a crucial limitation for a GaAs/AlAs heterostructure, for which QWs can be grown with wider width.
\end{thebibliography}
\end{document}